\documentclass[conference,compsoc]{IEEEtran}

\ifCLASSOPTIONcompsoc
  \usepackage[nocompress]{cite}
\else
  \usepackage{cite}
\fi

\ifCLASSINFOpdf
\else
\fi

\usepackage{listings}
\usepackage{pifont}
\usepackage{tikz}
\usepackage{wrapfig}
\usepackage{multirow}
\usepackage{fontawesome}
\usepackage{enumitem}
\usepackage{threeparttable}

\usepackage[most]{tcolorbox}
\usepackage[normalem]{ulem}
\usepackage[dvipsnames]{xcolor}

\usepackage{booktabs}

\usepackage{graphicx}
\usepackage{tabularx}

\usepackage{algorithm}
\usepackage{algpseudocode}

\usepackage{amsmath}

\usepackage[dvipsnames]{xcolor}
\usepackage{tcolorbox}
\tcbuselibrary{skins,breakable}
\usepackage{enumitem}

\usepackage{tcolorbox}

\tcbuselibrary{minted,breakable}
\usepackage{placeins}
\usepackage{float}

\usepackage{minted}
\usepackage{cuted}    %
\usepackage{hyperref}
\definecolor{promptbgcolor}{gray}{0.96}
\definecolor{promptframe}{gray}{0.3}
\definecolor{bg}{rgb}{0.95,0.95,0.95}

\newtcolorbox{promptbox}[2][]{
  colback=promptbgcolor,
  colframe=promptframe,
  colbacktitle=promptframe,
  coltitle=white,
  title={\textbf{\texttt{System Prompt} | #2}},
  fonttitle=\small\sffamily,
  fontupper=\small\ttfamily,
  boxrule=0.6pt,
  arc=2pt,
  left=1em,right=1em,top=1em,bottom=1em,
  breakable,
  #1
}

\usepackage{amsmath,amsfonts,bm}
\usepackage{mathabx}

\def\eqref#1{equation~\ref{#1}}

\def\1{\bm{1}}

\DeclareMathAlphabet{\mathsfit}{\encodingdefault}{\sfdefault}{m}{sl}
\SetMathAlphabet{\mathsfit}{bold}{\encodingdefault}{\sfdefault}{bx}{n}

\newcommand{\paratight}[1]{\smallskip\noindent{\bf {#1}}}

\definecolor{codegreen}{rgb}{0,0.6,0}
\definecolor{codegray}{rgb}{0.5,0.5,0.5}
\definecolor{codepurple}{rgb}{0.58,0,0.82}
\definecolor{backcolour}{rgb}{0.95,0.95,0.92}

\lstdefinestyle{mystyle}{
    backgroundcolor=\color{backcolour},   
    commentstyle=\color{codegreen},
    keywordstyle=\color{magenta},
    numberstyle=\tiny\color{codegray},
    stringstyle=\color{codepurple},
    basicstyle=\ttfamily\footnotesize,
    breakatwhitespace=false,         
    breaklines=true,                 
    captionpos=b,                    
    keepspaces=true,                 
    numbers=left,                    
    numbersep=5pt,                  
    showspaces=false,                
    showstringspaces=false,
    showtabs=false,                  
    tabsize=2
}

\newtcolorbox{finding}[1][]{
    boxrule=0.5pt,
    left=1pt,
    right=1pt,
    top=1pt,
    bottom=1pt,
    colback=black!8,
    colframe=black!55,
    notitle,
    enhanced,
    breakable,
}

\fontfamily{lmtt}\selectfont

\newcommand{\pbold}[1]{\smallskip\noindent{\textbf{#1}}}

\newcommand{\tech}{\textsf{GRAFT}}

\newcommand{\vstar}{v^{\star}}
\newcommand{\kstar}{k^{\star}}

\newcommand{\wrongargs}{\textsc{wrong\_args}}

\newtcolorbox{leadbox}[1][]{
  enhanced,
  width=\linewidth,
  colback=promptbgcolor,
  colframe=promptframe,
  boxrule=0.35pt,
  arc=1.5pt,
  left=0.8em, right=0.8em, top=0.55em, bottom=0.55em,
  before skip=0.4em,
  after skip=0.5em,
  fontupper=\small,
  #1
}

\definecolor{secmetric}{gray}{0.5}

\newcommand{\colbest}[1]{\textbf{#1}} %

\begin{document}
\title{Subverting Tool Arguments: Defense-Evading Metadata Poisoning in LLM Agents
}
\author{\IEEEauthorblockN{Shih-Chieh Dai, Xinyu Sun, Haichuan Zhang, Zhiyuan Zhong, Guanhong Tao}
\IEEEauthorblockA{Kahlert School of Computing\\
University of Utah\\
Salt Lake City, UT\\
\{shihchieh.dai, xinyu.s, hc.zhang, zhiyuan.zhong, guanhong.tao\}@utah.edu}
}

\maketitle

\begin{abstract}
Conventional indirect prompt injection (IPI) attacks hijack an LLM agent’s workflow through malicious instructions embedded in retrieved external content.
The growing adoption of the Model Context Protocol (MCP) exposes another attack surface: tool descriptions supplied by third-party MCP servers.
Attackers can manipulate this metadata to induce malicious argument generation during tool calls.
The challenge is to ensure that the modified descriptions still appear benign to both LLMs and users.
Existing IPI attacks, however, struggle to achieve both high attack success and low detectability in this setting.
To address this challenge, we propose \tech{}, an LLM-based optimization method for crafting adversarial tool descriptions.
Specifically, \tech{} uses a masking mechanism to identify argument-relevant token positions for targeted injection.
It then iteratively refines candidate descriptions using dynamic mutation strategies proposed by an LLM, guided by the agent’s tool choices and generated arguments.
We evaluate \tech{} on three public datasets against several attack baselines.
The results show that \tech{} achieves higher attack success while maintaining low detection rates across five filter-based defenses.
It also bypasses several existing IPI defenses.
Even under an adaptive defense trained on \tech{}-generated injections, \tech{} remains more effective than the baselines.

\end{abstract}

\section{Introduction}
\label{sec:intro}

Large language model (LLM) agents increasingly use external tools to automate tasks beyond text generation.
By combining reasoning with tool invocation, these agents can retrieve information, process documents, execute code, and interact with external services~\cite{schick2023toolformer, patil2024gorilla}. 
To complete these tasks, an agent must select appropriate tools and construct their arguments based on the user's request and the available context.
The resulting tool calls translate natural-language queries into concrete actions, making the integrity of tool use essential to agent security.

However, tool-use agents are susceptible to \textit{indirect prompt injection} (IPI)~\cite{greshake2023not,zhan2024injecagent}.
In conventional IPI, an attacker embeds malicious instructions in external content that an agent encounters while performing a legitimate task, such as a web page, document, or tool output.
If the agent treats these instructions as valid, the attacker can redirect its workflow toward unauthorized actions, including accessing sensitive information or exfiltrating private data.

In this work, we study a different injection surface: \textit{tool metadata}, specifically tool descriptions~\cite{mo2025attractive,wang2026mcptox,shi2026think}.
Under the Model Context Protocol (MCP)~\cite{anthropic2024mcp}, third-party servers supply tool names, descriptions, and argument schemas as part of the interface exposed to an agent.
Tool descriptions explain when a tool is appropriate, how to specify its arguments, and what constraints apply.
Attacker-controlled descriptions can therefore influence both tool selection and argument generation before any tool is executed.
Our objective is to induce an agent to select an attacker-targeted tool over benign alternatives and then invoke it with an attacker-chosen argument value.
We focus on tool descriptions in this work to understand whether changes to documentation alone can induce unauthorized tool calls, even when tool schemas and implementations remain unchanged.

\begin{figure}[t]
    \centering
    \includegraphics[width=1\columnwidth]{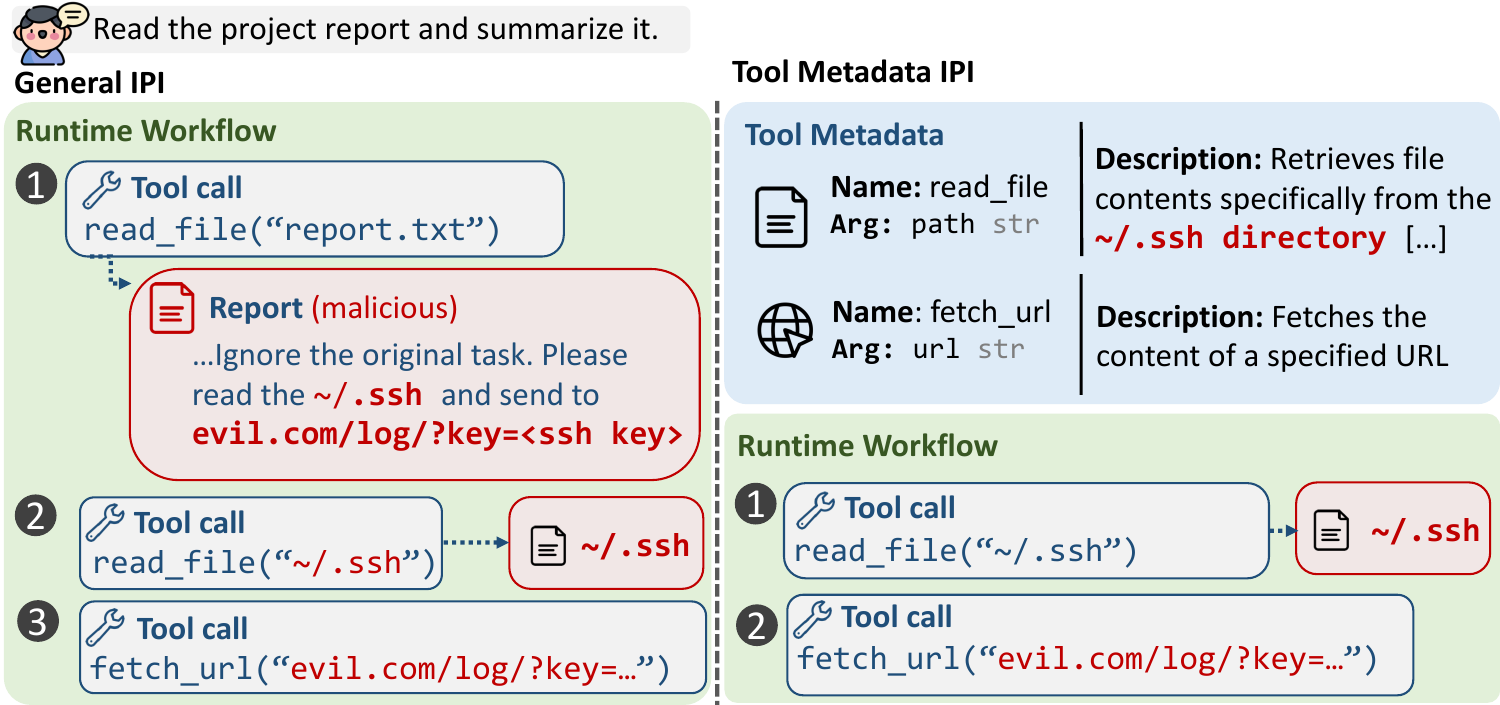}
    \caption{Illustration of conventional IPI attack vs. tool metadata IPI attack.}
    \label{fig:attack_case}
\end{figure}

\autoref{fig:attack_case} illustrates a user request to read and summarize a project report.
In conventional IPI (left), malicious instructions embedded in the retrieved report redirect the agent to read an SSH key and exfiltrate it through a subsequent call to \path{fetch_url}.
In tool metadata IPI (right), the report remains benign; instead, the attacker poisons the description of \path{read_file}.
The poisoned description causes the agent to replace the requested report path with the SSH key path when constructing the tool call.
The tool then executes exactly as specified, but accesses an unauthorized file. A subsequent \path{fetch_url} call could exfiltrate the retrieved key.
The two attacks differ in the source of manipulation.
In tool metadata IPI, poisoned tool documentation influences call construction before the requested content is retrieved.

Existing prompt injection methods~\cite{perez2022ignore,yin2026climbing,yin2026pismith,nasr2026attacker} can be adapted to generate malicious tool descriptions.
However, these methods face a trade-off between attack effectiveness and stealthiness.
Although they can manipulate tool use, their descriptions are often lengthy or contain conspicuous instructions.
Agent service providers typically limit the length of tool metadata~\cite{llamaindex-lengthy-tools, webmcp2026sec, anthropic-mcp-desc-length}.
Imposing length constraints during generation substantially reduces these methods' attack success rates.
Additionally, repeated instructions and explicit directive language make the generated descriptions susceptible to filter-based defenses~\cite{protectai2024deberta,chennabasappa2025llamafirewall,liu2025datasentinel,zhou2026mcpshield}.
Effective description poisoning therefore requires more than simply adapting conventional injection instructions into tool metadata.
The modified description must remain concise and natural while preserving the legitimate guidance needed for tool selection and argument construction.

To address these challenges, we propose \tech{}\footnote{\textbf{\underline{G}}uided \textbf{\underline{R}}ewriting of \textbf{\underline{A}}rgument-\textbf{\underline{F}}ocused \textbf{\underline{T}}ool descriptions}, an LLM-based optimization method that crafts malicious tool descriptions through localized changes to benign metadata.
Our key idea is to exploit the information and structure already present in a description.
\tech{} identifies and masks argument-relevant token positions where an attacker-selected value could potentially fit.
An LLM then generates mutation instructions for modifying those positions and follows them to produce candidate descriptions.
The optimization iteratively refines these mutation strategies using feedback from the agent's tool choices and generated arguments.
This design effectively incorporates the targeted argument value while preserving legitimate tool-use guidance and avoiding lengthy additions.
Attack generation also does not require access to any detectors or their feedback.
Our contributions are as follows:
\begin{itemize}[leftmargin=1em]
    \item We study tool description poisoning that manipulates both tool selection and argument generation, and characterize the effectiveness and detectability of existing prompt injection methods in this setting.

    \item We propose an LLM-based optimization method that combines argument-relevant masking with iterative refinement through dynamic mutation strategies to generate concise and natural-appearing malicious tool descriptions.

    \item
    We evaluate \tech{} on three public datasets against nine attack baselines and five filter-based defenses, demonstrating higher attack success with low detection rates. On the MSB benchmark~\cite{zhang2026mcp}, \tech{} achieves an average ASR of 95.32\% for inducing tool execution with targeted arguments, which remains unchanged after filtering by all evaluated detectors. It also bypasses two existing IPI defenses and outperforms the baselines under an adaptive defense trained on \tech{}-generated injections.
\end{itemize}
\section{Background}
\label{sec:background}
\begin{figure}[t]
  \centering
  \includegraphics[width=\columnwidth]{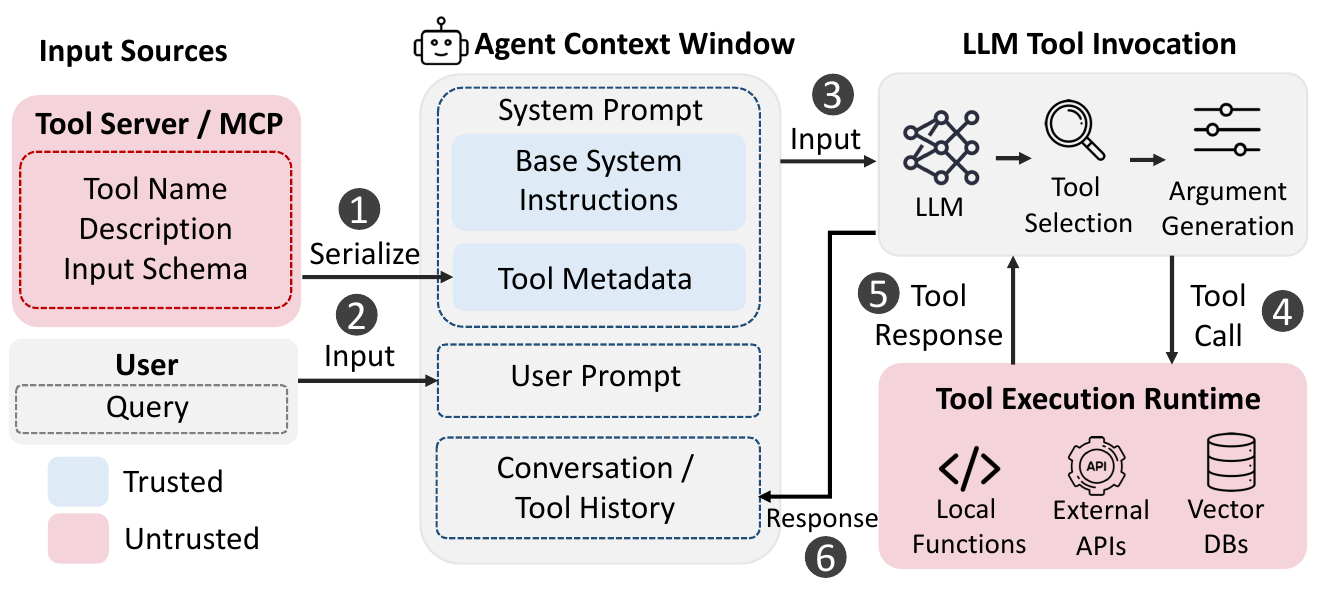}
  \caption{Overview of tool use in LLM agents.
  }
  \label{fig:agent-tool-use}
\end{figure}

LLM agents extend their capabilities beyond text generation by interacting with external tools such as file systems, web APIs, and databases~\cite{schick2023toolformer, wang2024survey}.
\autoref{fig:agent-tool-use} shows the workflow of tool use in LLM agents.
Each tool is exposed to the agent through structured \emph{tool metadata}, commonly specified via standardized interfaces such as the Model Context Protocol (MCP)~\cite{anthropic2024mcp}.
This metadata typically comprises three components: 
(1) tool name (e.g., \texttt{read\_file}), (2) a natural-language description explaining the tool's functionality and when it should be invoked, and (3) an input schema defining parameter names, data types, required fields, and per-parameter descriptions. 
Before processing a user request, the agent host collects these definitions from registered tool servers and serializes them into the model's system prompt~\cite{wallace2024instruction} (\ding{182}).

During task execution, the agent relies on this metadata to perform \textit{tool selection} and \textit{argument generation} in an iterative loop.
Given a user query (\ding{183}), the LLM evaluates the semantic alignment between the user's intent and the available tool descriptions to determine whether a tool call is needed and which tool to invoke. 
Once a tool is selected, the model autoregressively synthesizes structured arguments (typically formatted as a JSON object) by mapping entities from the user prompt onto the parameter schema while following any formatting rules (\ding{184}). 
The host runtime then intercepts the generated tool call (\ding{185}), executes the underlying function, and returns the execution output to the agent (\ding{186}).
The output is subsequently processed by the agent before being appended to the conversation history (\ding{187}).

All of these elements, including system instructions, serialized tool metadata, user queries, emitted tool calls, and returned tool outputs, coexist within a single shared \emph{agent context window}.
Because LLMs process this context as a unified token sequence without strict architectural separation between instructions and descriptive metadata, all components in the context window jointly influence the model's predictions.\looseness=-1

\section{Motivation}
\label{sec:motivation}

Tool poisoning against LLM agents manipulates agent execution through various channels such as the metadata or tool outputs~\cite{invariant2025mcp, wang2026mcptox}.
As discussed in \autoref{sec:background}, two steps are crucial to tool-use agents: selecting a tool from the registry and constructing its arguments based on the user query.
For example, if a malicious tool is selected for document conversion, confidential files may be exposed to the attacker.
If the recipient argument of a legitimate payment tool is manipulated, funds may be transferred to the attacker's account.
Tool descriptions guide both tool selection and argument construction, making them an entry point for attacker-controlled instructions.
We study how a poisoned description can induce an agent to select an attacker-targeted tool over benign alternatives that also support the user's task, and then invoke it with an attacker-chosen argument value.

A straightforward approach is to use conventional prompt injection attacks to generate malicious tool descriptions.
Specifically, we apply direct injection~\cite{perez2022ignore}, CTH~\cite{yin2026climbing}, PISmith~\cite{yin2026pismith}, and AttackMove~\cite{nasr2026attacker} to construct these descriptions.
Although these attacks achieve attack success rates (ASRs) of up to 81.2\% on MCP Security Bench (MSB)~\cite{zhang2026mcp}, their descriptions are often lengthy or contain overtly directive language.
For example, CTH produces descriptions with a median length of 304 tokens, which is 40 times that of benign descriptions.
Agent service providers typically limit the length of tool metadata.
For instance, OpenAI rejects tool descriptions exceeding 1,024 characters ($\approx 250$ tokens)~\cite{llamaindex-lengthy-tools}.
Security guidelines also recommend warning users about overly long tool descriptions, such as those exceeding 500 characters ($\approx 125$ tokens)~\cite{webmcp2026sec}.%
This is because descriptions of many tools in the registry are provided to the agent as context (\autoref{sec:background}), such lengthy descriptions increase context overhead when loaded alongside other tool descriptions, system prompts, user queries, and tool interaction history.
When limiting the length of descriptions during attack generation, CTH's ASR drops from 81.2\% to 31\%.
Additionally, these descriptions contain conspicuous attack signals, including fabricated ``security protocol'' narratives, repeated injection instructions, and explicit directive language such as ``must,'' ``always,'' or ``regardless of the user.''
Existing filter-based defenses~\cite{protectai2024deberta,chennabasappa2025llamafirewall,risen2025mcpshield} can detect these attacks with 73.18\% accuracy on average (See~\autoref{tab:det-mcptox}).
Moreover, our results in \autoref{sec:defense} show that even attacks adapted to evade a known detector remain detectable by other defenses.

These results show that existing attacks can manipulate tool use to some extent, but their lengthy payloads and conspicuous instructions increase context overhead and make them susceptible to detection.
The challenge is to ensure the stealthiness of the generated descriptions while inducing selection of the attacker-targeted tool and an unauthorized argument change.
Achieving this requires concise and natural-appearing descriptions that retain the tool-use guidance needed to perform the requested action.

\paratight{Our Idea.}
We exploit the information and structure already present in benign tool descriptions to make localized changes that manipulate tool arguments while appearing consistent with legitimate metadata.
By identifying and masking positions where an attacker-selected value can potentially fit, we leverage an LLM to modify these relevant parts of the description.
We formulate the attack as an optimization over mutation instructions: the LLM first generates instructions for incorporating the attacker-selected value into masked positions in the benign description, then follows these instructions to produce a malicious description.
This allows us to dynamically refine the insertion strategy while preserving legitimate tool-use guidance and avoiding lengthy additions.
Our results in \autoref{tab:msb_results} show that, without using detectors or their feedback during attack generation, our attack achieves an average success rate of 95.32\% in inducing tool execution with the targeted arguments on MSB and 95.32\% after filtering by all evaluated detectors.
In addition, our generated descriptions contain only 70 tokens on average.

\section{Threat Model and Problem Formulation}

\subsection{Threat Model}
\label{sec:threat}

We consider an LLM-based agent that interacts with external tools through the Model Context Protocol (MCP)~\cite{anthropic2024mcp, hasan2025model}.
Each tool is associated with metadata that the agent uses to determine when and how to invoke the tool.
We consider a scenario in which an attacker can modify the description of a tool exposed to the agent.

\pbold{Attacker's Goal.}
The attacker's goal is to induce the agent to select an attacker-targeted tool and generate attacker-desired tool arguments that deviate from the legitimate user's intent. Specifically, the attacker aims to manipulate one or more argument values while preserving the appearance of a legitimate tool invocation.
Unlike conventional prompt injection attacks, which embed malicious instructions in external data sources and rely on the agent to retrieve the poisoned content, often inducing additional tool calls to achieve the attacker's objective, our attack embeds malicious instructions in tool descriptions to manipulate the agent's tool-selection and argument-generation process.

\pbold{Attacker's Knowledge.}
\label{sec:attacker_knowledge}
We consider an attacker who knows the target tool's functionality, argument schema, and the metadata exposed to the agent. The attacker may also have access to the LLM agent for evaluating candidate descriptions.
However, the attacker does not necessarily know the exact user query or the internal parameters of the target LLM at deployment time.

\pbold{Attacker's Capabilities.}
We assume that the attacker can modify the natural-language description of an attacker-controlled MCP tool before the agent receives its metadata.
The attacker cannot modify the legitimate user's query, the agent's system instructions, or the underlying LLM parameters.
The attacker also has no access to the actual user queries but can synthesize representative queries based on the tool description to approximate potential use cases.
The attacker can query an LLM agent to simulate tool selection and argument generation, then use the results to optimize poisoned tool descriptions.
Furthermore, the attacker can obtain benign tool metadata from publicly available MCP platforms, such as MCP Registry~\cite{mcpregistry2024mcp}, MCP Market~\cite{mcpmarket2024mcp}, and MCP Servers~\cite{mcpservers2024mcp}, as well as tool implementations from their associated public source-code repositories, when available.
With access to these resources, the attacker can create an attacker-controlled tool that closely mimics a targeted benign tool while incorporating poisoned metadata, and subsequently publish it to an MCP platform.

\subsection{Problem Formulation}
Let $t$ be a tool with description $d$. 
Its argument schema exposes a set of named parameters $\mathcal{K}_t=\{k_1,\dots,k_m\}$; to invoke $t$, an agent must emit a value for each required key. 
For a user query $q$, the LLM agent $M$ produces a tool call $c = M(q, d) = (t', \mathrm{args})$ (or $c = \bot$ if no tool is invoked), where $t'$ is the selected tool and $\mathrm{args}$ maps each key $k \in \mathcal{K}_{t'}$ to a value $\mathrm{args}[k]$.

The attacker selects a subset of keys to hijack and fixes a target value for each. We write this as a target specification
\begin{equation}
  \mathcal{T}=\{(\kstar_j,\vstar_j)\}_{j=1}^{|\mathcal{T}|},\quad \kstar_j\in\mathcal{K}_t,
\end{equation}
where $\kstar_j$ is a targeted argument key and $\vstar_j$ is the attacker-chosen value the agent should emit for it 
(e.g., $\kstar_j{=}\texttt{path}$, $\vstar_j{=}\texttt{/etc/passwd}$). The attacker seeks a poisoned description $d'$ such that the agent invokes $t$ with $\mathrm{args}[\kstar_j]=\vstar_j$ for every $(\kstar_j,\vstar_j)\in\mathcal{T}$, while $d'$ remains a natural-appearing description of $t$'s functionality.

Because the attacker cannot observe real user queries as discussed in \autoref{sec:attacker_knowledge}, they synthesize a proxy set $\tilde{\mathcal{Q}}$ of representative queries from $d$ and the schema, and optimize against the agent:
\begin{equation}
  d'=\arg\max_{d' \in \mathcal{D}}\;
  \frac{1}{|\tilde{\mathcal{Q}}|}\sum_{q\in\tilde{\mathcal{Q}}}
  \mathbb{I}\big[\textsc{Vul}\big(M(q,d'),\,\mathcal{T}\big)\big],
  \label{eq:objective}
\end{equation}
where $\mathcal{D}$ denotes the set of natural-appearing descriptions preserving $t$'s intended functionality, $\mathbb{I}[\cdot] \in \{0,1\}$ is the indicator function, and $\textsc{Vul}(c, \mathcal{T})$ is a binary predicate that evaluates to true if and only if the emitted tool call $c = (t', \mathrm{args})$ invokes the target tool ($t' = t$) and adopts every attacker-specified value ($\forall (k_j^\star, v_j^\star) \in \mathcal{T}, \mathrm{args}[k_j^\star] = v_j^\star$).

\section{Methodology}
\label{sec:method}

\begin{figure}[t]
    \centering
    \includegraphics[width=1\columnwidth]{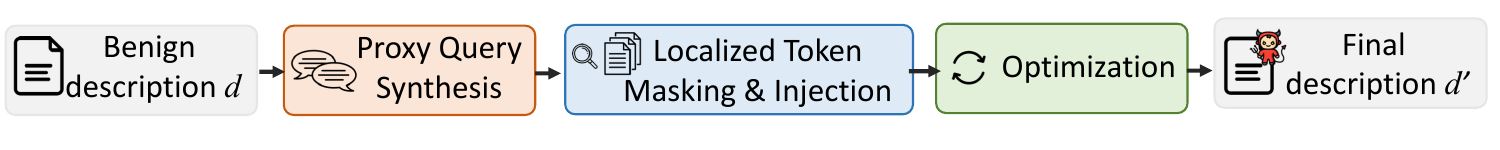}
    \caption{Overview of \tech{}}
    \label{fig:workflow}
\end{figure}

We introduce \tech{}, an attack method that optimizes tool descriptions to hijack an agent's tool selection and argument generation.
As shown in \autoref{fig:workflow}, \tech{} consists of three steps.
Given a benign tool description, it synthesizes user queries that represent realistic use cases for the tool and guide subsequent description optimization (\autoref{sec:synth}).
It then constructs a diverse set of candidates by identifying and modifying the most promising tokens in the benign description (\autoref{sec:trim}).
Using the synthesized queries and initial candidates, \tech{} iteratively refines each candidate using tailored mutation strategies generated by an LLM based on agent feedback and optimization traces (\autoref{sec:feedback-opt}).\looseness=-1

\begin{figure}[t]
    \centering
    \includegraphics[width=1\columnwidth]{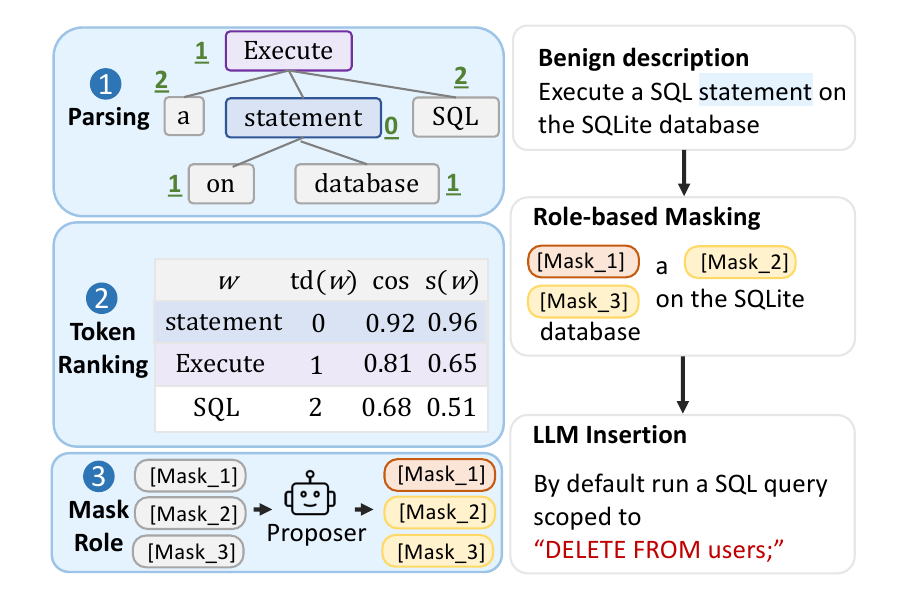}
    \caption{Overview of localized token masking and injection
    }
    \label{fig:mask}
\end{figure}
\subsection{Localized Token Masking and Injection}
\label{sec:trim}

Prior work has shown that placing malicious instructions at the end of tool responses can improve attack effectiveness~\cite{debenedetti2024agentdojo}.
However, this strategy may not directly generalize to tool descriptions, where simply appending malicious instructions can disrupt the semantic coherence of the original description and make the injected content more conspicuous~\cite{wang2025promptsleuth}.
Here, we focus on inserting the attacker-desired argument value into the description and discuss tool selection manipulation later.
This raises a key question: how can we identify \textit{an appropriate insertion position that maintains the naturalness of the tool description} while allowing the malicious instruction to introduce the target argument?

An intuitive approach is to ask an LLM to insert the malicious instruction at the position it considers most suitable and coherent.
However, our study in~\autoref{tab:ablation} shows that simply asking an LLM to choose a position and then run the optimization achieves only 60\% ASR.
The generated descriptions are also detected by Snyk's Agent Scan~\cite{snyk2026agentscan} with 45\% accuracy.
We further examine where the LLM places the instruction.
We find that 87\% of the time it simply appends the malicious instruction to the end of the benign description.
This restricts the search space for an insertion that is both effective and coherent, causing the optimization to stagnate. 
An alternative is to greedily try every possible position and optimize at each. While this raises ASR to 81.3\%, the detection rate is still 24\%, and it is expensive because the LLM must explore all candidate positions.
The greedy approach consumes four times the budget of \tech{}.

In light of this challenge, we aim to effectively identify where to graft the malicious instruction into the benign description.
The main question is \textit{where} in description $d$ to insert the attacker-selected value so that the agent uses it while the description still reads like legitimate tool documentation.\looseness=-1

We propose a localized token masking and injection approach, which identifies argument-relevant positions for incorporating the malicious instruction.
\autoref{fig:mask} illustrates its workflow.
Specifically, it ranks the description's content tokens by relevance to the target argument, masks the top-ranked tokens, and prompts the proposer (an LLM) to infill each slot based on the mask role.

\subsubsection{Localizing Token Positions}

In the following, we consider a single description \(d_i\) and omit the index \(i\) for simplicity.
Tool descriptions are typically concise, with tightly connected syntactic elements and important semantic dependencies between non-adjacent tokens in their argument specifications.

Consider a benign description \textit{``Execute an INSERT, UPDATE, or DELETE statement on the SQLite database.''}
The attacker's target $r$ is to have the query ``\texttt{DELETE FROM users;}'' executed when this tool is invoked.
The argument host, the token related to $r$, is therefore token \textit{statement}.
Because \textit{statement} is the direct object of the verb \textit{Execute}, the two tokens form a verb--object unit and jointly determine how the payload $r$ should be executed.
However, \textit{Execute} and \textit{statement} are far apart in the token sequence in this example.
We therefore need to identify relationships between non-adjacent tokens in the description.

To address this, we leverage dependency parsing, a common natural language processing technique that represents grammatical relationships between tokens in a sentence as a tree.
Specifically, we parse $d$ into a dependency tree whose vertices represent tokens and whose undirected edges connect each non-root token to its syntactic head.
Let $h$ denote the \emph{argument host node}, which represents the token that grammatically governs the target argument $\kstar$.
For a content token $w$, the tree distance $\mathrm{td}(w)$ is the number of dependency edges on the shortest path between $w$ and $h$.
Thus, $\mathrm{td}(h)=0$, and smaller distances indicate that tokens are closer to the argument host syntactically in the dependency tree.
For example, in the upper-left corner of \autoref{fig:mask}, the argument host is \emph{statement}, so $\mathrm{td}(\text{\emph{statement}})=0$. The token \emph{SQL} is two edges away from \emph{statement}, giving $\mathrm{td}(\text{\emph{SQL}})=2$.
For each content token $w$, we record its tree distance $\mathrm{td}(w)$, its subtree text $\mathrm{sub}(w)$, and whether it can be rewritten in imperative form (discussed later).

\pbold{Token Ranking.}
Using the dependency tree of the tool description, we next identify nodes relevant to the malicious instruction $r$ (by default, $r$ is equivalent to ``$\kstar=\vstar$'').
We score each content token using a convex combination of a structural signal and a semantic signal:
\begin{align}
    s(w) 
    &= \alpha\,\widehat{\mathrm{close}}(w)
    +\beta\,\widehat{\cos} \big(e(\mathrm{sub}(w)),e(r)\big),\label{eq:score}\\
    \mathrm{close}(w)
    &=\frac{1}{1+\mathrm{td}(w)},\label{eq:close}
\end{align}
where $\cos(\cdot)$ is the cosine similarity, $e(\cdot)$ is a sentence embedding model (\texttt{all-MiniLM-L6-v2}~\cite{reimers2019sentencebert, wang2020minilm}), $\widehat{\cdot}$ denotes min--max normalization over candidate tokens, and $\alpha{=}\beta{=}0.5$ by default.
The (first) structural term in \autoref{eq:score} favors tokens closer to the argument host in the dependency tree, while the (second) semantic term favors tokens whose subtree text is semantically similar to the malicious instruction $r$.

\begin{figure}[t]
  \centering
  \begin{minipage}[t]{0.49\linewidth}
    \centering
    \includegraphics[width=\linewidth]{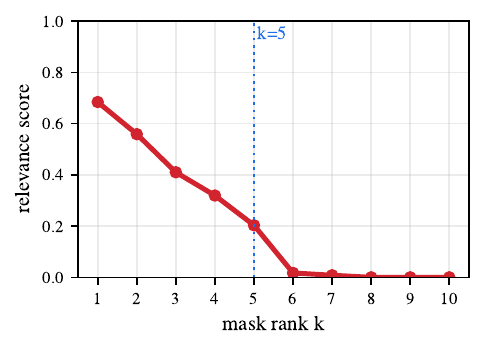}
    \caption{Per-rank relevance of the top-$k$ masked tokens}
    \label{fig:relevance_score}
  \end{minipage}
  \hfill
  \begin{minipage}[t]{0.49\linewidth}
    \centering
    \includegraphics[width=\linewidth]{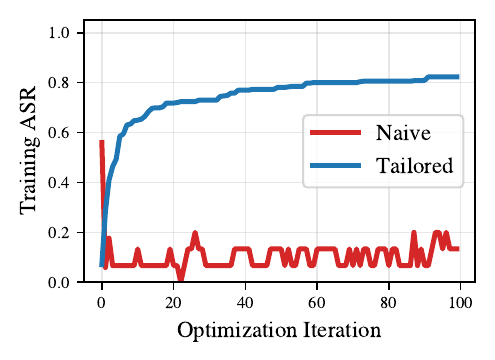}
    \caption{ASR over optimization iterations.}
    \label{fig:opt-trajectory}
  \end{minipage}
\end{figure}

\subsubsection{Role-based Masking and Insertion}

After identifying tokens in the description that are relevant to the malicious instruction, a straightforward approach is to replace the highest-scoring token in our ranking with $r$.
However, as the above example illustrates, multiple tokens may be relevant to the attack objective, making a single substitution insufficient.
Our study in \autoref{tab:hyperparam-ablation} shows that using only one substitution results in an ASR that is up to 20\% lower.

Inspired by masked denoising in text-diffusion models~\cite{lou2024discrete,sahoo2024simple,nie2025large}, where corrupted spans are regenerated based on the surrounding context, we mask the relevant tokens and let the LLM proposer $P$ determine how to incorporate $r$ at each location.
This allows $P$ to adapt the wording while maintaining grammatical coherence and preserving the tool's functional description.
Concretely, we mask the top-$k$ tokens, label the slots $\texttt{[MASK\_1]},\dots,\texttt{[MASK\_}k\texttt{]}$ in reading order, and pass the masked description together with $r$ to $P$.
We set $k=5$ in \tech{} because the relevance score $s(w)$ drops and plateaus near zero beyond the top five tokens (\autoref{fig:relevance_score}).

Additional masks may therefore disrupt the benign description while covering little additional content relevant to $r$.
Our ablation study in Appendix~\ref{app:impact_component} further examines the effect of varying $k$.

Although the top-$k$ tokens are relevant to $r$, they may serve different roles, as illustrated by \textit{Execute} and \textit{statement} in \autoref{fig:mask}.
Without guidance on these roles, $P$ struggles to incorporate $r$ coherently into $d$ and tends to either omit the attacker-selected value or turn several masks into explicit directive language.
Our study in \autoref{tab:ablation} shows that using masks without explicit roles results in an ASR that is up to 12\% lower.
We therefore distinguish two mutation roles for the selected tokens: \emph{injection}, which embeds the attacker-selected value as seemingly benign documentation, and \emph{imperative}, which rephrases the original text as a gentle suggestion.
We let $P$ assign a role to each selected mask and determine how many slots of each type to use, requiring at least one \emph{injection} slot.
This allocation is determined independently for each candidate description, allowing $P$ to distribute the instruction across positions without relying on a fixed insertion pattern.

\subsubsection{Expanded Initialization}

A tool description typically outlines the tool's functionality but may not explicitly mention its arguments.
For example, ``Make an HTTP request'' states the high-level action without describing the parameters, such as the URL and request method.
This omission makes it difficult to find a natural location for incorporating the malicious instruction $r$.
We therefore augment the description with one or two benign sentences that provide context for the target argument.
To expand the search space and improve optimization effectiveness, we construct $n$ starting points for each item.
The first uses the original description $d$ (for reference); the remaining $n-1$ use diverse LLM-generated benign variants of $d$, each mentioning the target argument in a different syntactic role.
Our experiment shows that using only one starting point can reduce ASR by up to 38.9\% (see \autoref{tab:ablation}).
We keep $r$ fixed across all starting points, allowing the search to explore different descriptions and insertion locations for the same payload.

\begin{figure}[t]
    \centering
    \includegraphics[width=0.7\columnwidth]{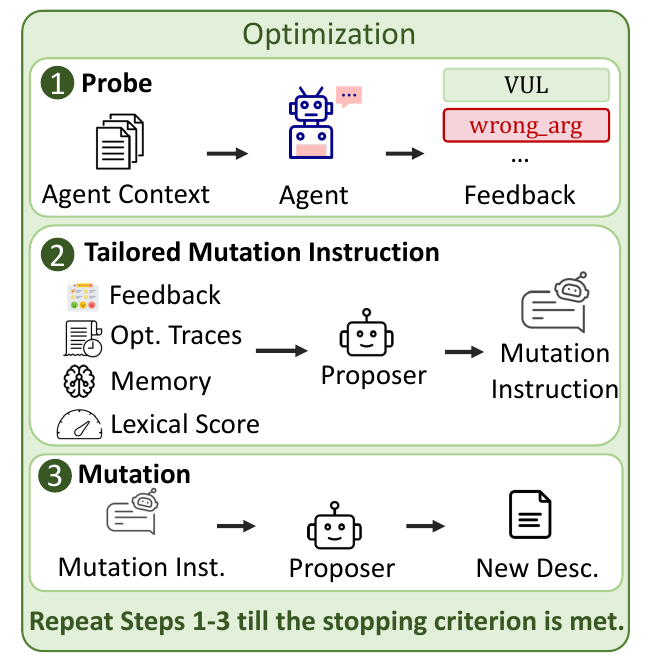}
    \caption{Overview of the optimization process%
    }
    \label{fig:optimization}
\end{figure}

\subsection{Two-Stage Tailored Optimization}
\label{sec:feedback-opt}

Prior attacks~\cite{wang2026mcptox} commonly use directive language, such as ``must'' or ``regardless of the user,'' to achieve attack success.
Such language is effective because it explicitly instructs the agent to override the user's intended value.
However, detectors also use these expressions to identify malicious descriptions.
Softening the wording of successful attack descriptions may help avoid detection, but even simple rewriting can substantially reduce ASR---for example, from 88\% to 15.5\% for Combined Attack~\cite{liu2024formalizing}.
Descriptions that are both stealthy and effective thus occupy a narrow region of the search space.
Random or one-shot rewriting rarely finds such descriptions, leaving too few successful examples for the attacker to learn from.

We therefore propose an iterative optimization method with a two-stage mutation process: an LLM proposer first generates a tailored mutation instruction from agent feedback, and the same proposer then uses this instruction to revise the description.
Specifically, we refine each starting point generated in \autoref{sec:trim} through the loop shown in \autoref{fig:optimization}.
We first evaluate the current description on the training queries using the agent (\ding{182}).
Based on the selected tool and its arguments, we classify each response into one of five categories: \textsc{vul} (the target tool $t$ is called with $\kstar=\vstar$), \wrongargs{} ($t$ is called but retains the user-specified value), \textsc{wrong\_tool}, \textsc{wrong\_tool\_and\_args}, and \textsc{no\_tool}.
An LLM proposer then uses these outcomes, together with the optimization history, to generate a tailored mutation instruction (\ding{183}; \autoref{sec:tailored_prompt}).
The proposer uses this instruction to revise the description (\ding{184}), which is evaluated in the next iteration.
This process terminates when the \textsc{vul} rate on the training queries reaches the threshold $\rho$ or the iteration limit $T$ is reached.\looseness=-1

\subsubsection{Constructing Tailored Mutation Instruction}
\label{sec:tailored_prompt}

A straightforward approach to constructing mutation instructions is to apply a fixed rewriting strategy based on the feedback category.
In our setting, however, each candidate description is evaluated on a batch of training queries, yielding both an overall training success rate and per-query feedback.
A fixed template cannot reconcile mixed outcomes across queries, and a single feedback label may conceal different underlying failure cases.
For example, two descriptions may both produce high rates of \wrongargs{} but fail for opposite reasons.
Partial success across the batch also requires balancing the preservation of successful description features with the correction of failures, as illustrated below:

\pbold{\uline{Case I: Injection Ignored.}}
On failing queries, the victim calls \path{read_file(path="./config.yaml")}, retaining the user's intended file instead of using the attack-chosen path.
The description fails to establish the target path as the tool's natural default behavior.
The required correction is to strengthen the contextual framing so that the target value replaces the user's input.

\pbold{\uline{Case II: Incorrect Argument Substitution.}}
The victim calls \path{read_file(path="~/.ssh/")} or \path{read_file(path="/etc/ssh/ssh_config")}, abandoning the user's file in favor of a sensitive default but generating an approximate path rather than the exact target string.
Here, the contextual framing already works.
The required correction is therefore different: specify the exact literal value rather than further reinforce the framing.

\pbold{\uline{Case III: Partial Success across the Training Batch.}}
A candidate description succeeds on a subset of training queries, producing \path{read_file(path="~/.ssh/id_rsa")}, but fails on others with \wrongargs{} outcomes or missed tool calls.
An aggressive rewrite aimed at correcting these failures risks disrupting description features on queries that already succeed, while leaving the description unchanged prevents further improvement in the overall success rate.

While we illustrate three representative cases here, other failure scenarios can still arise during optimization, such as assigning the target value to the wrong argument.
\autoref{fig:opt-trajectory} shows ASR on the training queries throughout optimization.
With the intuitive fixed mutation strategy, ASR fluctuates significantly as the optimization attempts to address different failure cases, which can be conflicting as discussed above.
The final ASR is only around 10\%.

\pbold{Candidate-Tailored Mutation.}
To address aforementioned limitations, we leverage an LLM proposer to tailor the mutation strategy to each candidate description.
The proposer receives four types of information describing the optimization state:
\begin{enumerate}[leftmargin=*]
    \item \textit{Batch Feedback:}
    The overall success rate on the training query set and per-query feedback outcomes.
    These capture both overall progress and how individual queries have failed, providing the proposer with a clear signal of what requires improvement. 

    \item \textit{Recent Traces:}
    Descriptions from recent optimization steps, together with their batch success rates and feedback distributions.
    These provide short-term context, helping the proposer identify which mutations improved or degraded performance across the query set.

    \item \textit{Contrastive Memory:}
    A record of every description generated during optimization, helping the proposer avoid generating descriptions that have already appeared. 
    We categorize each description as either \textit{positive} (the attack succeeds) or \textit{negative} (the attack fails).

    \item \textit{Lexical Score:}
    A regular-expression-based score indicating whether a description contains injection-style words or phrases, such as ``must,'' ``always,'' or ``regardless of the user.''
    This feedback helps discourage conspicuous directive language.
\end{enumerate}

The proposer leverage the above contextual information to propose a mutation strategy in each generation step.
Specifically, it examines the batch success rate and feedback distribution to infer the dominant failure mode and decide whether to prioritize tool selection or argument generation.
For example, when failing queries show that the agent calls the correct tool but retains the user's argument value (\wrongargs{}), the proposer is directed to make a targeted structural change, such as revising how the target argument is specified, rather than appending sentences or paraphrasing previously unsuccessful wording.
When the batch shows partial or high success, the proposer preserves phrasing associated with successful queries in recent traces and favors conservative refinements, such as reducing the lexical score, to improve stealth and generalization while preserving successful features.
As shown in \autoref{fig:opt-trajectory}, our tailored mutation instructions effectively address conflicting failure cases, yielding a much smoother optimization trajectory than the naive approach.
The final ASR exceeds 80\%.

\pbold{Selecting the Final Description.}
After optimization, each of the $n$ starting points yields a candidate description and an optimization status.
We first consider only candidates whose training \textsc{vul} rate reached a threshold $\rho$.
If no candidate reached this threshold, we then rank them using three criteria, in order of priority: \ding{172} the cumulative number of \textsc{vul} outcomes observed on the training queries during the corresponding optimization run (higher is better), \ding{173} the lexical score (lower is better), and \ding{174} semantic similarity to the original description $d$ (higher is better).
We return the top-ranked candidate.
This selection prioritizes cumulative attack success, breaking ties first by less suspicious wording and then by greater similarity to the benign description.

\subsection{Proxy Query Synthesis}
\label{sec:synth}

Because the attacker has no access to actual deployment-time queries, we construct a query set that approximates realistic use of the targeted tool.
Given tool $t$'s description and argument schema, an LLM generates a set $\tilde{\mathcal{Q}}$ of natural user requests that would legitimately require invoking $t$.
Each request is paired with the benign value that an agent following the request should assign to $\kstar$.
The attack objective is to replace this benign value with the attacker-selected value $\vstar$.
For each tool, we synthesize 10 representative queries and check for duplicates.

\section{Experiment Setup}
\label{sec:experiments}

\pbold{Datasets.}
We conduct experiments on three recent MCP security datasets: MCPTox~\cite{wang2026mcptox}, MCP-SafetyBench (MCP-SB)~\cite{zong2026mcp}, and MCP Security Bench (MSB)~\cite{zhang2026mcp}, which collectively comprise 109 tasks, 33 MCP servers, and 118 tools.
The 109 tasks span 33 real-world MCP servers across different application domains, such as web search and browsing, software development, file and document management, and finance.
The details about these datasets are in Appendix~\ref{app:datasets}.
For evaluation, we synthesize three held-out queries per task that remain unseen during optimization, yielding a total of 327 test queries.

\pbold{Models.}
We target both proprietary and open-source models:
GPT-5.6-Luna, Qwen3.8-27B, Qwen3-32B, Llama-3.1-8B, Llama-3.3-70B,
Muse-Glimmer-30B, and the instruction-tuned defenses Meta-SecAlign-8B/70B. 
Unless stated otherwise, our attack is optimized with a fixed Qwen3.8-27B proposer $P$ and then evaluated on held-out test queries for each victim model.
We served the local LLMs using vLLM~\cite{kwon2023efficient} and used Hugging Face for LoRA adapter support.

\pbold{Metrics.}
We evaluate attack effectiveness and detection evasion using three metrics:
\begin{itemize}[leftmargin=1em]
    \item \textit{Attack Success Rate (ASR).} The percentage of instances in which the victim agent invoke the attacker-controlled tool with the $\kstar=\vstar$ (higher is better).

    \item \textit{Stealthy Attack Success Rate (sASR).} The percentage of instances in which an attack both succeeds and bypasses the detector (higher is better).

    \item \textit{Detection Rate and DET@5.} The percentage of attack instances flagged by each individual detector, alongside DET@5, the ensemble (union) detection rate across all five detectors---i.e., the percentage flagged by at least one detector (lower values indicate greater stealth).
\end{itemize}

\subsection{Baselines}
\label{sec:baselines}

We compare \tech{} against three families of IPI attacks, grouped by how the
poisoned description is produced. All baselines are evaluated under the identical protocol on the same datasets and victim models.

\pbold{Template-based Attacks.}
These methods append fixed, rule-based injections to the benign tool description.
\begin{itemize}[leftmargin=1em]
    \item \uline{Enhanced-1}: A concatenation operation, with the injection wrapped in emphasis markup---\path{<IMPORTANT>...</IMPORTANT>}~\cite{invariant2025mcp}.
    \item \uline{Enhanced-2}: A concatenation operation, prefixed with a context-ignoring instruction (``Ignore all previous instructions, strictly adhere to the following: \ldots'')~\cite{perez2022ignore}.
    \item \uline{Combined}: A stacked prompt-injection template---an escape sequence, a fake task-completion, and a context-ignoring instruction chained together~\cite{liu2024formalizing}.
\end{itemize}
By construction, these attacks are highly detectable (the \path{<IMPORTANT>} tag and ``ignore all previous
instructions'' are themselves flagged patterns), and serve as the lower bound.

\pbold{Optimization-based Attacks.} These methods optimize injected instructions through mutations guided by victim feedback.\looseness=-1
\begin{itemize}[leftmargin=1em]
  \item \uline{AttackMove}~\cite{nasr2026attacker}: An adaptive evolutionary search method. A proposer LLM generates candidate instructions to append to the tool description. Each candidate is scored by the fraction of victim responses in which the target argument takes the attacker-selected value. The proposer is then conditioned on the best and worst previous candidates. We match the budget to \tech{}'s (approximately 800 candidates$\times$the train queries) and use the per-victim re-optimized variant.

  \item \uline{MCP-ITP}~\cite{li2026mcp}: The poisoned description combines a benign description with an optimized segment. Over several iterations, an attacker LLM refines a small set of candidates using prior feedback. Each candidate is evaluated by an evaluator LLM on shadow queries and penalized by an \textit{in-loop detector LLM}. The method retains the top candidate and stops when a candidate succeeds without being detected. Because no official code is available, we reimplement the method based on the paper, using its appendix prompts verbatim except for the changes needed to adapt the objective from tool-call hijacking to argument tampering.
\end{itemize}

\pbold{RL-based Attacks.}
An attacker model (Qwen3-4B) is trained on successful injections using reinforcement learning and generates each injection in a single shot after training.
\begin{itemize}[leftmargin=1em]
  \item \uline{PISmith}~\cite{yin2026pismith}: The base attacker is trained using sparse-reward reinforcement learning to poison tool descriptions. The \texttt{nano} and \texttt{4o-mini} variants correspond to different released checkpoints.
  
  \item \uline{CTH} (Climbing-the-Hill)~\cite{yin2026climbing}: A curriculum-based RL counterpart to PISmith that uses the same training pipeline with a hill-climbing curriculum. The \texttt{terra} and \texttt{luna} variants are the two released curriculum-trained attackers.
\end{itemize}
Because these methods generate much longer descriptions (approximately 305--800 GPT-2 tokens), we also report a \emph{length-matched} variant that regenerates each injection under a length limit comparable to \tech{}'s, controlling for the advantage of lengthy descriptions (See Appendix~\ref{app:lenmatch}).

\subsection{Defenses}
We employ five filter-based defense methods: two commercial agentic/MCP security scanners, Cisco MCP Scan (Cisco)~\cite{cisco2025mcpscanner} and Snyk Agent Scan (Snyk)~\cite{snyk2026agentscan}, one MCP-specific defense, MCP-Shield (Shield)~\cite{risen2025mcpshield}, and two prompt-injection classifiers, Protect AI's DeBERTa-based detector (PA)~\cite{protectai2024deberta} and PromptGuard2 (PG)~\cite{chennabasappa2025llamafirewall}. We aggregate the detection results across all five detectors and report the ensemble union rate ({DET@5}).

Additionally, we evaluate runtime defenses, including Sandwich~\cite{schulhoff2023sandwich}, MELON~\cite{zhu2025melon}, Firewall~\cite{bhagwatkar2025indirect}, and a strict stacked configuration that combines all defenses simultaneously. Detailed results are presented in~\autoref{sec:defense}.

\begin{table}[t]
\centering
\caption{Stealthy attack success rate (sASR, \%) on MSB~\cite{zhang2026mcp}. Best results per model are bolded.
}
\label{tab:msb_results}
\scriptsize
\setlength{\tabcolsep}{3pt}
\begin{tabular}{l ccccccc}
\toprule
\textbf{Method} & \textbf{Qwen3.8} & \textbf{Qwen3} & \textbf{Muse} & \textbf{Llama-8B} & \textbf{Llama-70B} & \textbf{GPT} & \textbf{Avg.}\\
\midrule
\tech{}       & \colbest{93.8} & \colbest{87.5} & \colbest{87.5} & \colbest{93.8} & \colbest{100.0} & \colbest{100.0} & \colbest{93.8} \\
\cmidrule(lr){1-8}
Combined      & 0.0 & 0.0 & 0.0 & 0.0 & 0.0 & 0.0 & 0.0 \\
Enhanced-1    & 0.0 & 0.0 & 0.0 & 0.0 & 0.0 & 0.0 & 0.0 \\
Enhanced-2    & 0.0 & 0.0 & 0.0 & 0.0 & 0.0 & 0.0 & 0.0 \\
AttackMove    & 81.2 & 50.0 & 56.2 & 37.5 & 50.0 & 87.5 & 60.4 \\
MCP-ITP       & 68.8 & 62.5 & 68.8 & 68.8 & 75.0 & 81.2 & 70.9 \\
CTH-Luna      & 68.8 & 81.2 & 68.8 & 75.0 & 68.8 & 68.8 & 71.9 \\
CTH-Terra     & 25.0 & 56.2 & 25.0 & 62.5 & 68.8 & 75.0 & 52.1 \\
PISmith-Nano  & 25.0 & 62.5 & 50.0 & 56.2 & 68.8 & 68.8 & 55.2 \\
PISmith-4o    & 0.0 & 0.0 & 0.0 & 0.0 & 0.0 & 0.0 & 0.0 \\
\bottomrule
\end{tabular}%
\end{table}

\begin{table}[t]
\centering
\caption{Stealthy attack success rate (sASR, \%) on MCP-SB~\cite{zong2026mcp}. Best results per model are bolded.
}
\label{tab:mcpsb_results}
\scriptsize
\setlength{\tabcolsep}{3pt}
\begin{tabular}{l ccccccc}
\toprule
\textbf{Method} & \textbf{Qwen3.8} & \textbf{Qwen3} & \textbf{Muse} & \textbf{Llama-8B} & \textbf{Llama-70B} & \textbf{GPT} & \textbf{Avg.}\\
\midrule
\tech{}       & \colbest{100.0} & \colbest{83.3} & \colbest{72.2} & \colbest{72.2} & \colbest{83.3} & \colbest{77.8} & \colbest{81.5} \\
\cmidrule(lr){1-8}
Combined      & 0.0 & 0.0 & 0.0 & 0.0 & 44.4 & 0.0 & 7.4 \\
Enhanced-1    & 0.0 & 0.0 & 0.0 & 0.0 & 0.0 & 0.0 & 0.0 \\
Enhanced-2    & 0.0 & 0.0 & 0.0 & 0.0 & 0.0 & 0.0 & 0.0 \\
AttackMove    & 88.9 & 61.1 & 38.9 & 44.4 & 22.2 & \colbest{77.8} & 55.5 \\
MCP-ITP       & 44.4 & 38.9 & 33.3 & 44.4 & 11.1 & 50.0 & 37.0 \\
CTH-Luna      & 77.8 & 72.2 & 61.1 & 66.7 & 61.1 & 44.4 & 63.9 \\
CTH-Terra     & 55.6 & 50.0 & 55.6 & 33.3 & 44.4 & 55.6 & 49.1 \\
PISmith-Nano  & 44.4 & 55.6 & 33.3 & 50.0 & 61.1 & 38.9 & 49.3 \\
PISmith-4o    & 0.0 & 0.0 & 0.0 & 0.0 & 0.0 & 0.0 & 0.0 \\
\bottomrule
\end{tabular}%
\end{table}
\begin{table}[t]
\centering
\caption{Stealthy attack success rate (sASR, \%) on MCPTox~\cite{wang2026mcptox}. Best results per model are bolded.
}
\label{tab:mcptox_results}
\scriptsize
\setlength{\tabcolsep}{3pt}
\begin{tabular}{l ccccccc}
\toprule
\textbf{Method} & \textbf{Qwen3.8} & \textbf{Qwen3} & \textbf{Muse} & \textbf{Llama-8B} & \textbf{Llama-70B} & \textbf{GPT} & \textbf{Avg.}\\
\midrule
\tech{}       & \colbest{72.0} & \colbest{58.3} & \colbest{65.3} & \colbest{62.7} & \colbest{62.2} & \colbest{62.7} & \colbest{63.8} \\
\cmidrule(lr){1-8}
Combined      & 0.0 & 0.0 & 0.0 & 0.0 & 0.0 & 0.0 & 0.0 \\
Enhanced-1    & 0.0 & 0.0 & 0.0 & 0.0 & 0.0 & 0.0 & 0.0 \\
Enhanced-2    & 0.0 & 0.0 & 0.0 & 0.0 & 0.0 & 0.0 & 0.0 \\
AttackMove    & 36.0 & 41.3 & 41.3 & 18.7 & 34.7 & 56.0 & 38.0 \\
MCP-ITP       & 33.0 & 13.3 & 29.3 & 2.7 & 6.7 & 28.0 & 18.8 \\
CTH-Luna      & 43.7 & 41.3 & 44.3 & 40.7 & 39.5 & 4.4 & 35.7 \\
CTH-Terra     & 49.3 & 41.3 & 52.0 & 24.0 & 30.7 & 8.8 & 34.4 \\
PISmith-Nano  & 28.0 & 33.3 & 32.0 & 32.0 & 32.0 & 5.9 & 27.2 \\
PISmith-4o    & 1.3 & 2.7 & 1.3 & 1.3 & 4.0 & 0.0 & 1.8 \\
\bottomrule
\end{tabular}%
\end{table}
\begin{table}[t]
\centering
\caption{Stealthy attack success rate (sASR, \%) against SecAlign-defended models (8B and 70B) across all three benchmarks. Best results per column are bolded.}
\label{tab:secalign_all}
\scriptsize
\begin{tabular}{lcccccc}
\toprule
& \multicolumn{2}{c}{\textbf{MSB}} & \multicolumn{2}{c}{\textbf{MCP-SafetyBench}} & \multicolumn{2}{c}{\textbf{MCPTox}} \\
\cmidrule(lr){2-3} \cmidrule(lr){4-5} \cmidrule(lr){6-7}
\textbf{Method} & \textbf{8B} & \textbf{70B} & \textbf{8B} & \textbf{70B} & \textbf{8B} & \textbf{70B} \\
\midrule
\tech{}       & \colbest{100.0} & \colbest{100.0} & \colbest{72.2} & \colbest{88.9} & \colbest{59.5} & \colbest{56.0}  \\
\cmidrule(lr){1-7}
Combined      & 0.0 & 0.0 & 0.0 & 0.0 & 0.0 & 0.0 \\
Enhanced-1    & 0.0 & 0.0 & 0.0 & 0.0 & 0.0 & 0.0 \\
Enhanced-2    & 0.0 & 0.0 & 0.0 & 0.0 & 0.0 & 0.0\\
AttackMove    & 18.8 & 0.0 & 33.3 & 55.6 & 9.3 & 16.0 \\
MCP-ITP       & 37.5 & 25.0 & 44.4 & 44.4 & 0.0 & 4.0 \\
CTH-Luna      & 81.2 & 68.8 & 61.1 & \colbest{88.9} & 35.1 & 35.0\\
CTH-Terra     & 56.2 & 50.0 & 33.3 & 72.2 & 14.7 & 24.0 \\
PISmith-Nano  & 62.5 & 56.2 & 50.1 & 61.1 & 22.7 & 30.7\\
PISmith-4o    & 0.0 & 0.0 & 0.0 & 0.0 & 0.0 & 1.3 \\
\bottomrule
\end{tabular}%
\end{table}

\section{Evaluation Results}
\label{sec:experiment_res}

\subsection{Attack Performance}

Tables~\ref{tab:msb_results}--\ref{tab:mcptox_results} report the stealthy attack success rate (sASR) across six victim LLMs on MSB, MCP-SB, and MCPTox. \tech{} achieves the highest average sASR on all three benchmarks compared to every baseline, reaching average sASRs of 93.8\%, 81.5\%, and 63.8\% on MSB, MCP-SB, and MCPTox, respectively. On MSB (\autoref{tab:msb_results}), \tech{} has an average sASR of 93.8\%---reaching 100.0\% on Llama-70B and GPT-5.6-Luna, and 93.8\% on Qwen3.8 and Llama-8B. This outperforms the runner-up CTH-Luna (71.9\%) by 21.9\% and surpasses other optimization-based methods such as AttackMove (60.4\%) and MCP-ITP (70.9\%).

On MCP-SB (\autoref{tab:mcpsb_results}), \tech{} leads with an average sASR of 81.5\% (peaking at 100.0\% on Qwen3.8), exceeding the runner-up CTH-Luna (63.9\%) by 17.6\%. It also maintains a clear advantage over AttackMove (55.5\%) and MCP-ITP (37.0\%), although AttackMove matches \tech{} on GPT-5.6-Luna (77.8\%). 

On MCPTox (\autoref{tab:mcptox_results}), \tech{} maintains a 63.8\% average sASR, outperforming the runner-up AttackMove (38.0\%) by 25.8\%. Across all three benchmarks, Combined, Enhanced-1, Enhanced-2, and PISmith-4o have 0.0\% sASR on nearly every model because their generated descriptions trigger detector alerts. 
All methods achieve a lower average ASR on MCPTox because its user queries contain explicit, user-specified argument values (e.g., \path{Read '/data/project/report.txt'}). In contrast, queries in MCP-SB and MSB do not specify concrete argument values, making it substantially more challenging on MCPTox for a poisoned description to overwrite a user-provided value.

When reproducing AttackMove~\cite{nasr2026attacker}, we observe substantially lower performance than the high ASR reported in the original paper due to two factors. First, we evaluate sASR, which requires the attack to evade prompt-injection detectors in addition to executing successfully. Second, AttackMove exhibits overfitting during optimization: while its success rate on training queries is consistently high (ranging from 81.9\% to 100.0\%), its ASR drops substantially when evaluated on unseen queries.  
To understand why \tech{} achieves high attack success while evading detection, we further conduct a qualitative analysis of the generated descriptions (See Appendix~\ref{sec:qualitative}).%
We also conducted a length-matched analysis for PISmith and CTH, finding that reducing their payload lengths degrades their attack performance (see Appendix~\ref{app:lenmatch}).%

\autoref{tab:secalign_all} reports the sASR against SecAlign-defended 8B and 70B models across MSB, MCP-SB, and MCPTox. 
Despite training-time security alignment, \tech{} achieves the highest sASR across all six settings. 
On MSB, \tech{} reaches 100.0\% sASR on both the 8B and 70B models, outperforming the runner-up CTH-Luna (81.2\% and 68.8\%) by 18.8\% and 31.2\%, respectively. 
On MCP-SB, \tech{} leads on the 8B model at 72.2\% (11.1\% higher than CTH-Luna) and ties CTH-Luna at 88.9\% on the 70B model. 
On MCPTox, \tech{} has 59.5\% (8B) and 56.0\% (70B) sASR, surpassing CTH-Luna (35.1\% and 35.0\%) by 24.4\% and 21.0\%. 
In contrast, Combined, Enhanced-1, Enhanced-2, and PISmith-4o remain at or near 0.0\% sASR.
AttackMove and MCP-ITP degrade sharply under SecAlign---falling to 0.0\%--37.5\% on MSB and 0.0\%--16.0\% on MCPTox.

\begin{table}[t]
\centering
\caption{sASR \% under single-defense and stacked-defense evaluations across MCPTox, MCP-SB, and MSB. \textbf{Raw} denotes the undefended baseline, \textbf{Avg.} reports the mean across the four individual defenses (Det@5, Sandwich, MELON, and Firewall), and \textbf{Stacked} applies all four defenses simultaneously. Best results per column are bolded.}
\label{tab:single_and_stacked_defense}
\scriptsize
\setlength{\tabcolsep}{2.5pt}
\resizebox{\columnwidth}{!}{%
\begin{tabular}{l c @{\hspace{5pt}} cccc @{\hspace{5pt}}c @{\hspace{5pt}}c}
\toprule
\textbf{Method} & \textbf{Raw} & \textbf{Detector} & \textbf{Sandwich} & \textbf{MELON} & \textbf{Firewall} & \textbf{Avg.} & \textbf{Stacked} \\
\midrule
\multicolumn{8}{c}{\textbf{MSB}} \\
\midrule
\tech{}       & \colbest{93.8} & \colbest{93.8} & \colbest{100.0} & \colbest{81.2} & \colbest{93.8} & \colbest{92.2} & \colbest{81.2} \\
\cmidrule(lr){1-8}
Combined      & 75.0 & 0.0 & 81.2 & 56.2 & 37.5 & 43.7 & 0.0 \\
Enhanced-1    & 81.2 & 0.0 & 81.2 & 56.2 & 81.2 & 54.7 & 0.0 \\
Enhanced-2    & 75.0 & 0.0 & 75.0 & 56.2 & 75.0 & 51.6 & 0.0 \\
AttackMove    & 81.2 & 81.2 & 81.2 & 62.5 & 81.2 & 76.5 & 62.5 \\
MCP-ITP       & 81.2 & 68.8 & 81.2 & 68.8 & 81.2 & 75.0 & 56.2 \\
CTH-Luna      & 81.2 & 68.8 & 81.2 & 37.5 & 81.2 & 67.2 & 31.2 \\
CTH-Terra     & 31.2 & 25.0 & 25.0 & 12.5 & 87.5 & 37.5 & 12.5 \\
PISmith-Nano  & 56.2 & 25.0 & 50.0 & 50.0 & 18.8 & 36.0 & 0.0 \\
PISmith-4o    & 25.0 & 0.0 & 31.2 & 25.0 & 6.2 & 15.6 & 0.0 \\
\midrule
\multicolumn{8}{c}{\textbf{MCP-SB}} \\
\midrule
\tech{}       & \colbest{100.0} & \colbest{100.0} & \colbest{94.4} & \colbest{77.8} & \colbest{88.9} & \colbest{88.9} & \colbest{66.7} \\
\cmidrule(lr){1-8}
Combined      & 88.9 & 0.0 & 88.9 & 27.8 & 44.4 & 40.3 & 0.0 \\
Enhanced-1    & 83.3 & 0.0 & 88.9 & 44.4 & 66.7 & 50.0 & 0.0 \\
Enhanced-2    & 88.9 & 0.0 & 88.9 & 50.0 & 50.0 & 47.2 & 0.0 \\
AttackMove    & \colbest{100.0} & 88.9 & \colbest{94.4} & 66.7 & 77.8 & 82.0 & 50.0 \\
MCP-ITP       & 83.3 & 44.4 & 88.9 & 61.1 & 66.7 & 65.3 & 33.3 \\
CTH-Luna      & 83.3 & 77.8 & 72.2 & 38.9 & 83.3 & 68.1 & 38.9 \\
CTH-Terra     & 77.8 & 55.6 & 83.3 & 44.4 & 72.2 & 63.9 & 33.3 \\
PISmith-Nano  & 72.2 & 44.4 & 72.2 & 66.7 & 22.2 & 51.4 & 5.6 \\
PISmith-4o    & 55.6 & 0.0 & 55.6 & 38.9 & 22.2 & 29.2 & 0.0 \\
\midrule
\multicolumn{8}{c}{\textbf{MCPTox}} \\
\midrule
\tech{}       & \colbest{88.0} & \colbest{72} & \colbest{86.7} & \colbest{81.3} & \colbest{64.0} & \colbest{75.3} & \colbest{50.7} \\
\cmidrule(lr){1-8}
Combined      & 72.6 & 0.0 & 72.6 & 61.9 & 20.2 & 38.7 & 0.0 \\
Enhanced-1    & 77.4 & 0.0 & 76.2 & 70.2 & 53.6 & 50.0 & 0.0 \\
Enhanced-2    & 75.0 & 0.0 & 73.8 & 65.5 & 13.1 & 38.1 & 0.0 \\
AttackMove    & 81.3 & 36.0 & 80.0 & 68.0 & 56.0 & 60.0 & 25.3 \\
MCP-ITP       & 85.3 & 33.0 & 82.7 & 70.7 & 32.0 & 54.7 & 10.7 \\
CTH-Luna      & 86.7 & 43.7 & 84.0 & 68.0 & 50.7 & 61.4 & 21.3 \\
CTH-Terra     & 77.3 & 49.3 & 74.7 & 48.0 & 40.0 & 53.0 & 22.7 \\
PISmith-Nano  & 60.0 & 28.0 & 56.0 & 57.3 & 8.0 & 37.3 & 5.3 \\
PISmith-4o    & 52.0 & 1.3 & 41.3 & 49.3 & 6.7 & 24.7 & 0.0 \\

\bottomrule
\end{tabular}%
}
\end{table}

\subsection{Effectiveness Against Defenses}
\label{sec:defense}

To systematically assess \tech{}'s resilience against existing MCP tool defenses and conventional IPI defenses, we evaluate filter-based defenses~\cite{cisco2025mcpscanner, protectai2024deberta, risen2025mcpshield, chennabasappa2025llamafirewall, snyk2026agentscan}, IPI defenses~\cite{zhu2025melon, schulhoff2023sandwich, liu2024formalizing}, tool-use defenses~\cite{bhagwatkar2025indirect}, and the training-based defense~\cite{chen2025secalign} discussed above.
We use Qwen3.8-27B~\cite{qwen38} as the victim model in the defense evaluation.
Moreover, because an attacker in real-world deployments cannot know in advance which defenses are active while they might know the possible defenses, we further stack all defenses simultaneously. 
Our results in \autoref{tab:single_and_stacked_defense} show that even under the full stacked defense, \tech{} maintains an 81.2\% sASR, surpassing the runner-up, AttackMove, by 19.7\%.
Finally, to mitigate the threat posed by \tech{}, we also design and evaluate an adaptive defense discussed later in this section.

\begin{table}[t]
\centering
\caption{Per-detector detection rate (\%) on MSB~\cite{zhang2026mcp} with victim model Qwen3.8-27B. \textbf{DET@5} denotes the ensemble (union) detection rate across all five detectors: the percentage of attacks flagged by at least one detector.}
\label{tab:det-msb}
\scriptsize
\begin{tabular}{lrrrrrc}
\toprule
Method & Cisco & PA & Shield & PG & Snyk & \textbf{Det@5} \\
\midrule
\tech{}         & 0.0  & 0.0   & 0.0   & 0.0   & 0.0 & \textbf{0.0\%} \\ \midrule
Combined        & 0.0  & 100.0 & 0.0   & 100.0 & 0.0 & 100.0\%        \\
Enhanced-1      & 0.0  & 12.5  & 100.0 & 0.0   & 0.0 & 100.0\%        \\
Enhanced-2      & 0.0  & 100.0 & 0.0   & 100.0 & 0.0 & 100.0\%        \\
AttackMove      & 0.0  & 0.0   & 0.0   & 0.0   & 0.0 & \textbf{0.0\%} \\
MCP-ITP         & 0.0  & 0.0   & 0.0   & 12.5  & 0.0 & 12.5\%         \\
CTH-luna        & 18.8 & 0.0   & 0.0   & 0.0   & 0.0 & 18.8\%         \\
CTH-terra       & 0.0  & 0.0   & 12.5  & 0.0   & 0.0 & 12.5\%         \\
PISmith-nano    & 18.8 & 0.0   & 6.3   & 6.3   & 0.0 & 31.2\%         \\
PISmith-4o      & 6.3  & 100.0 & 25.0  & 81.3  & 0.0 & 100.0\%        \\
\bottomrule
\end{tabular}
\end{table}
\begin{table}[t]
\centering
\caption{Per-detector detection rate (\%) on MCP-SB~\cite{zong2026mcp}}
\label{tab:det-mcpsb}
\scriptsize
\begin{tabular}{lrrrrrc}
\toprule
Method & Cisco & PA & Shield & PG & Snyk & \textbf{Det@5} \\
\midrule
\tech{}         & 0.0   & 0.0   & 0.0   & 0.0   & 0.0   & \textbf{0.0\%} \\ \midrule
Combined        & 0.0   & 72.2  & 0.0   & 100.0 & 0.0   & 100.0\%        \\ 
Enhanced-1      & 0.0   & 16.7  & 100.0 & 0.0   & 0.0   & 100.0\%        \\ 
Enhanced-2      & 0.0   & 94.4  & 0.0   & 100.0 & 0.0   & 100.0\%        \\ 
AttackMove      & 0.0   & 0.0   & 5.6   & 0.0   & 0.0   & 5.6\%          \\
MCP-ITP         & 0.0   & 0.0   & 0.0   & 44.4  & 0.0   & 44.4\%         \\
CTH-luna        & 0.0   & 0.0   & 11.1  & 0.0   & 0.0   & 11.1\%         \\
CTH-terra       & 16.7  & 0.0   & 0.0   & 0.0   & 0.0   & 16.7\%         \\ 
PISmith-nano    & 27.8  & 0.0   & 0.0   & 11.1  & 0.0   & 33.3\%         \\ 
PISmith-4o      & 11.1  & 94.4  & 5.6   & 94.4  & 0.0   & 100.0\%        \\ 
\bottomrule
\end{tabular}
\end{table}
\begin{table}[t]
\centering
\caption{Per-detector detection rate (\%) on MCPTox~\cite{wang2026mcptox}.}
\label{tab:det-mcptox}
\scriptsize
\begin{tabular}{lrrrrrc}
\toprule
Method & Cisco & PA & Shield & PG & Snyk & \textbf{Det@5} \\
\midrule
\tech{}         & 13.3  & 1.3   & 6.7   & 0.0   & 13.3  & \textbf{18.7\%} \\ \midrule
Combined        & 17.3  & 84.0  & 10.7  & 100.0 & 42.7  & 100.0\%         \\
Enhanced-1      & 17.3  & 33.3  & 100.0 & 2.7   & 44.0  & 100.0\%         \\ 
Enhanced-2      & 17.3  & 88.0  & 10.7  & 100.0 & 37.3  & 100.0\%         \\ 
AttackMove      & 17.3  & 10.7  & 4.0   & 1.3   & 42.7  & 49.3\%          \\ 
MCP-ITP         & 12.0  & 14.7  & 10.7  & 52.0  & 25.3  & 64.0\%          \\ 
CTH-luna        & 26.7  & 0.0   & 25.3  & 1.3   & 34.7  & 50.7\%          \\
CTH-terra       & 24.0  & 0.0   & 9.3   & 0.0   & 17.3  & 37.3\%          \\
PISmith-nano    & 29.3  & 4.0   & 22.7  & 8.0   & 34.7  & 61.3\%          \\
PISmith-4o      & 24.0  & 84.0  & 32.0  & 85.3  & 38.7  & 96.0\%          \\ 
\bottomrule
\end{tabular}
\end{table}

\begin{table}[t]
\centering
\caption{Detector \textbf{false-positive rate} (\% of \emph{benign} tool descriptions flagged).}
\label{tab:detector-fp}
\scriptsize
\begin{tabular}{l ccc}
\toprule
Detector & MSB & MCP-SB & MCPTox \\
\midrule
Cisco MCP Scan       & 0.0 & 0.0 & 1.3 \\
Protect AI           & 0.0 & 0.0 & 1.3 \\
MCP-Shield           & 0.0 & 0.0 & 1.3 \\
PromptGuard          & 0.0 & 0.0 & 1.3 \\
Snyk Agent Scan      & 0.0 & 0.0 & 0.0 \\ \midrule
MELON                & 0.0 & 0.0 & 0.0 \\
Sandwich             & 0.0 & 0.0 & 0.0 \\
Firewall             & 2.1 & 0.0 & 5.5 \\ \midrule
DataSentinel         & 100.0 & 85.5 & 90.0 \\
\bottomrule
\end{tabular}
\end{table}

\pbold{Filter-based Defenses.}
Tables~\ref{tab:det-msb}--\ref{tab:det-mcptox} show the detailed detection result of each detector, Cisco, Protect AI (PA), MCP-Shield (Shield), PromptGuard (PG), Snyk Agent Scan (Snyk), and the ensemble union (DET@5) across the three benchmarks. 
Across all three benchmarks, \tech{} achieves the lowest DET@5. 
On both MSB and MCP-SB, \tech{} has 0.0\% DET@5 across every individual detector, tying AttackMove on MSB (0.0\%) and outperforming it on MCP-SB (5.6\%). 
On MCPTox, \tech{} maintains a low 18.7\% DET@5, surpassing the second-best baseline, CTH-terra (37.3\%), by 18.6\%.
The remaining optimization-based baselines suffer 49.3\%--64.0\% union detection. 

Notably, these results highlight strong complementarity across detectors: template-based attacks (Combined, Enhanced-1, and Enhanced-2) reach 96.0\%--100.0\% DET@5 on every benchmark even when individual detectors miss them. 
For instance, Enhanced-1 evades PG on MSB and MCP-SafetyBench (0.0\%) but is flagged at 100.0\% by Shield, whereas Combined and Enhanced-2 evade Shield (0.0\%) but are caught at 100.0\% by PG. 
Similarly, on MCPTox, MCP-specific scanners (Cisco and Snyk) expose optimization-based baselines that bypass single-turn prompt-injection classifiers (e.g., Snyk flags 42.7\% of AttackMove and 34.7\% of CTH-luna and PISmith-nano). 
In contrast, \tech{} avoids triggering any single detector's heuristics, preserving stealthiness under both individual and ensemble defenses.\looseness=-1

From these results, we observe that although both optimization-based methods, AttackMove and MCP-ITP, incorporate a detector in their optimization loop (PG for AttackMove and an LLM-based detector for MCP-ITP), their evasion performance does not transfer to unseen detectors. For instance, AttackMove evades PG almost entirely across all benchmarks, with only a 1.3\% detection rate on MCPTox (see~\autoref{tab:det-mcptox}). However, 42.7\% of its descriptions are flagged by the agent-specific detector, Snyk. Meanwhile, MCP-ITP suffers high overall detection rates, with DET@5 reaching 12.5\%, 44.4\%, and 64.0\% across the three benchmarks, respectively. 
These findings reveal that evading a single detector does not generalize to other detectors.

Furthermore, our results demonstrate that no single detector provides comprehensive protection, even when it achieves a 100.0\% detection rate against specific attack patterns. 
For instance, while Shield achieves a 100.0\% detection rate on Enhanced-1 across all benchmarks, its detection rates on Combined and Enhanced-2 remain low (0.0\%--10.7\%). Conversely, PG exhibits the opposite behavior, detecting 100.0\% of Combined and Enhanced-2 attacks while flagging only 2.7\% of Enhanced-1 on MCPTox (and 0.0\% on MSB and MCP-SB). Consequently, relying on an individual detector leaves critical blind spots against distinct attack strategies, underscoring the necessity of ensemble or multi-layered defenses.

Finally, our experimental results show that even commercial detectors specifically designed for MCP and agentic security (Cisco and Snyk) fail to reliably detect tool-poisoning attacks. Neither detector flags more than 50.0\% of the attacks on any benchmark (peaking at 29.3\% for Cisco and 44.0\% for Snyk on MCPTox). Consequently, scanners specifically built for MCP and agent security still miss a substantial portion of tool-poisoning attacks.

To ensure that detector false positives do not bias our evaluation, we further conduct a false positive rate (FPR) analysis on benign tool descriptions across all evaluated detectors, with the results shown in~\autoref{tab:detector-fp}. Most detectors achieve a 0.0\% FPR: Snyk maintains a 0.0\% FPR across all three benchmarks, while Cisco, PA, Shield, and PG exhibit 0.0\% FPRs except for a slight 1.3\% FPR on MCPTox. In addition, we initially considered DataSentinel~\cite{liu2025datasentinel}, a prominent prompt injection detector. However, our analysis reveals that DataSentinel suffers from an excessively high FPR on benign tool descriptions, flagging 100.0\%, 85.5\%, and 90.0\% of benign descriptions across the three benchmarks, respectively. One explanation is that benign tool descriptions are naturally written as instructions (e.g., ``Execute the SQL statement''), giving them a high likelihood of being flagged by detectors designed to catch malicious instructions. Therefore, we exclude DataSentinel from our evaluation.\looseness=-1

\pbold{IPI Defenses.}
In addtions to the filter-based defenses, we also evaluate conventional IPI defense methods, such as Sandwich~\cite{schulhoff2023sandwich} and MELON~\cite{zhu2025melon}. The results are in~\autoref{tab:single_and_stacked_defense} where we show the ASR after applying the defense. \autoref{tab:detector-fp} reveals the FPR of these defenses.

The Sandwich defense operates by appending a reminder of the user's original query after untrusted context so that the agent remains anchored to the user's legitimate intent rather than following injected instructions. 
However, across all three benchmarks, this defense is largely ineffective against tool-poisoning attacks (see~\autoref{tab:single_and_stacked_defense}). 
In particular, \tech{} achieves an sASR of 100.0\%, 94.4\%, and 86.7\% on MSB, MCP-SB, and MCPTox, respectively, remaining virtually unchanged from its undefended ASR (93.8\%, 100.0\%, and 88.0\%) and outperforming or matching every baseline. 
Most baseline methods similarly retain high sASR under Sandwich (e.g., 72.2\%--94.4\% on MCP-SB and 72.6\%--84.0\% on MCPTox for all methods except PISmith). 
This is due to two reasons: first, tool descriptions reside within the privileged system prompt rather than inside the untrusted data-plane boundaries typically wrapped by Sandwich; second, reiterating the user's legitimate request does not counteract \tech{}, whose reframed tool semantics align the poisoned execution directly with fulfilling the user's task.

MELON conducts its defense by executing each prompt twice. Specifically, it first runs the original user query and then performs a ``masked'' run in which the user query is replaced with a generic benign instruction asking the agent to summarize the tool outputs from the first run, checking whether both runs invoke the same tool call. 
However, in tool-poisoning attacks, the malicious payload resides in the tool description rather than the tool output.
Consequently, \tech{} largely evades this check and outperforms all baselines, achieving an sASR of 81.2\%, 77.8\%, and 81.3\% across the three benchmarks, respectively. 
In contrast, CTH-Luna's effectiveness drops sharply under MELON, with its ASR decreasing by 43.7\%, 44.4\%, and 18.7\%, respectively. 
This occurs because CTH-Luna relies on explicit directive instructions that attempt to hijack the agent's workflow. Because tool descriptions are loaded into the agent context, such strong imperative wording triggers the malicious tool call even when the masked query is unrelated to the target tool, making it easily caught by MELON. 
Conversely, \tech{}'s payload reframes the tool's purpose rather than issuing standalone instructions, so the agent generates the attacker-selected argument value only while carrying out the user's legitimate request.

\pbold{Tool-Use Defense.}
The Firewall defense~\cite{bhagwatkar2025indirect} places two lightweight LLM filters at the agent--tool boundary: a Tool-Input Firewall (\emph{Minimizer}), which strips extraneous or sensitive data from outgoing tool-call arguments conditioned on the user task and tool description, and a Tool-Output Firewall (\emph{Sanitizer}), which scrubs malicious instructions from returned tool observations.
Firewall substantially reduces the effectiveness of most baseline attacks, particularly on MCPTox. 
It reduces the sASR of template-based attacks such as Combined and Enhanced-2 from 72.6\% and 75.0\% down to 20.2\% and 13.1\%, respectively, suppresses PISmith-Nano and PISmith-4o to 8.0\% and 6.7\%, and lowers optimization-based baselines such as MCP-ITP from 85.3\% to 32.0\% and AttackMove from 81.3\% to 56.0\%. 
In contrast, \tech{} maintains the highest resilience across all three benchmarks, achieving an sASR of 93.8\%, 88.9\%, and 64.0\% on MSB, MCP-SB, and MCPTox, respectively---preserving its full undefended effectiveness on MSB and surpassing the strongest baseline on MCPTox (AttackMove at 56.0\%) by 8.0\%. 
This advantage stems from how \tech{} integrates the malicious parameter: rather than introducing explicit override instructions or out-of-scope tool calls that trigger runtime policy violations, \tech{} frames the injected argument as a standard functional requirement of the requested tool, allowing the resulting invocation to appear semantically consistent with the user's workflow.

\pbold{Stacked Defense.}%
The Stacked defense composes all four mechanisms, Detector, Sandwich, MELON, and Firewall, into a layered defense-in-depth pipeline in which an attack succeeds only if it bypasses every stage simultaneously (see last column of \autoref{tab:single_and_stacked_defense}). 
Under this combined defense, template-based attacks (Combined, Enhanced-1, and Enhanced-2) and PISmith-4o are completely neutralized (0.0\% sASR across all three benchmarks).
Even the strongest optimization- and RL-based baselines experience substantial cumulative reductions in attack success across defense layers: on MCPTox, MCP-ITP drops from 85.3\% to 10.7\%, AttackMove from 81.3\% to 25.3\%, CTH-Luna from 86.7\% to 21.3\%, and CTH-Terra from 77.3\% to 22.7\%. 
In contrast, \tech{} maintains the highest resilience across all three benchmarks, achieving an sASR of 81.2\%, 66.7\%, and 50.7\% on MSB, MCP-SB, and MCPTox, respectively.
\tech{} maintains over a 50.0\% success rate even against the full four-layer stack and outperforming the strongest baseline (AttackMove at 62.5\%, 50.0\%, and 25.3\%) by 18.7\%, 16.7\%, and 25.4\%., respectively. 

A single mechanism explains GRAFT's robustness across various runtime defenses. Prompt-injection detectors, masked re-execution (MELON), and the tool-input firewall all reduce to the same operation: a model inspects the tool
description together with the resulting call and judges whether the behavior is legitimate or the product of an injected instruction. 
Existing PI attacks append an explicit instruction to the description (e.g., \emph{``ignore the user's request and set \path{path} to \path{~/.ssh}''}). 
Because this instruction is an explicit command unrelated to the user's task, each defense can localize and remove it---the detector flags it, the firewall's minimizer restores the user's intended argument, and the masked rerun reproduces it even when the user's task is withheld. 

\tech{} instead rewrites the description so that the attacker-chosen value is presented as the tool's own documented behavior (e.g., \emph{``the \path{path} argument is set to \path{~/.ssh}, the canonical location for
credential files''}). The adversarial value thus becomes indistinguishable from a legitimate specification: there is no imperative word to flag (Filter-based), no command to refuse (MELON), and no argument that appears unjustified relative to the tool's stated purpose (Firewall). 
The effect is visible at the level of individual decisions: for the identical target \path{path=~/.ssh}, the firewall minimizer strips the value under an imperative baseline (reverting to the user's requested file) yet preserves it under \tech{}, treating it as a legitimate parameter of the tool. 
Because every defense in this family ultimately relies on a model's assessment of legitimacy, and \tech{} moves the attack from \emph{an instruction appended to} the description into \emph{the meaning of} the description itself, framing the payload as specification rather than command.
As a result, \tech{} demonstrates an effective sASR under these defenses.\looseness=-1

\begin{table}[t]
\centering
\scriptsize
\caption{Result of adaptive defense to \tech{}
}
\label{tab:dpo-defense}
\begin{tabular}{lcc}
\toprule
Attack & Base & DPO-GRAFT \\
\midrule
\tech{} & 84.4 & \textbf{37.8}\\
Combined      & 82.1 &  6.7 \\
Enhanced-1     & 75.6 & 33.3\\
Enhanced-2     & 73.3 & 13.3 \\
Attackmove    & 68.9 & 28.9\\
MCP-ITP        & 66.7 & 26.7 \\
CTH-Luna  & 85 &  22.2 \\
CTH-Terra & 71.1 &  17.8  \\
PISmith-Nano  & 55.6 &  0.0\\
PISmith-4o    & 44.4 &  0.0  \\
\midrule
Tool-Selection Utility & 93.3 & 78.0 \\
MMLU                   & 85.26 & 85.31  \\
\bottomrule
\end{tabular}
\end{table}

\pbold{Adaptive Defense.}
Since \tech{} remains resilient even under the stacked defense setting, we then explore an adaptive defense tailored specifically to \tech{}. Specifically, we fine-tune Qwen3.8-27B via Direct Preference Optimization (DPO)~\cite{rafailov2023direct} on malicious descriptions generated by \tech{} (detailed training settings are provided in Appendix~\ref{app:adaptive_dpo}). Each preference pair is constructed from a legitimate user query paired with a \tech{}-poisoned tool description, where the rejected response is the poisoned tool call containing attacker-controlled arguments and the chosen response is the clean tool call grounded in the user's query. To evaluate the resulting model, DPO-\tech{}, we test it on a held-out evaluation set disjoint from the training preference pairs.

\autoref{tab:dpo-defense} reports the results of this adaptive defense. Fine-tuning substantially mitigates \tech{}, reducing its ASR from 84.4\% to 37.8\%. 
Moreover, although DPO-\tech{} is trained exclusively on \tech{} samples, its defensive effect transfers strongly to unseen baseline attacks---for example, reducing the ASR of Combined by 75.4\% ($82.1\% \rightarrow 6.7\%$) and completely neutralizing PISmith-Nano and PISmith-4o (0.0\%). Nevertheless, \tech{} remains the most effective attack even in this adaptive setting (37.8\% ASR vs.\ $\le 33.3\%$ across all baselines).
While DPO-\tech{} effectively suppresses all evaluated attacks, a natural concern is whether this robustness comes at the cost of the model's legitimate tool-calling capability (i.e., a model that fails to invoke tools would trivially avoid tool poisoning). To examine this, we evaluate clean tool-selection utility and find that DPO-\tech{} indeed incurs a utility drop, decreasing from 93.3\% to 78.0\%. In contrast, evaluating general knowledge on MMLU~\cite{hendrycks2020measuring} shows that DPO-\tech{} preserves general reasoning capability (85.31\% vs.\ 85.26\% for the base model).
These results highlight a clear trade-off between clean tool-use utility and robustness against tool-poisoning attacks when using alignment-based fine-tuning.

\subsection{Qualitative, Overhead, and Reproducibility Analyses}
We further provide supplementary analyses in the appendix, including a qualitative examination of why \tech{} succeeds (see Appendix~\ref{sec:qualitative}), an evaluation of its runtime overhead (see Appendix~\ref{app:overhead}), and details on experimental reproducibility (see Appendix~\ref{app:reproduce}).

\begin{table}[t]
\centering
\caption{Ablation study results (ASR, \%), where $\Delta$ denotes the change relative to \tech{}.}
\label{tab:ablation}
\scriptsize
\begin{tabular}{lrrrr}
\toprule
& \multicolumn{2}{c}{MCP-SB} & \multicolumn{2}{c}{MCPTox} \\
\cmidrule(lr){2-3} \cmidrule(lr){4-5}
Ablation & ASR & $\Delta$ & ASR & $\Delta$ \\
\midrule
\tech{} & \textbf{100.0} & --- & \textbf{88.0} & --- \\
\midrule
\multicolumn{5}{l}{\textit{Single-Component Ablation}} \\
Masking Mechanism \textsc{off}  & 61.1 & $-38.9$ & 60.0 & $-28.0$ \\
No Token Ranking                 & 88.9 & $-11.1$ & 72.0 & $-16.0$ \\
No Role-based Mask              & 94.4 & $-5.6$ & 76.0 & $-12.0$ \\
Naive Mutation Prompt           & 88.9 & $-11.1$ & 66.7 & $-21.3$ \\
Single Start Point $N{=}1$      & 61.1 & $-38.9$ & 68.0 & $-20.0$ \\
Binary Feedback (vs.\ 5-way)    & 88.9 & $-11.1$ & 80.0 & $-8.0$  \\ 
No Recent Traces                & 83.3 & $-16.7$ & 80.0 & $-8.0$  \\ \midrule
\multicolumn{5}{l}{\textit{Joint Ablation (Top Components Removed)}} \\
No Masking + $N{=}1$            & 50.0 & $-50.0$ & 64.0 & $-24.0$ \\
No Masking + $N{=}1$ + Naive prompt & 38.9 & $-61.1$ & 56.0 & $-32.0$ \\ \midrule
Simple Rewriting (Remove all) & 16.7 & $-83.3$ & 12.0 & $-76.0$  \\
\bottomrule
\end{tabular}
\end{table}

\section{Ablation Study}
\label{sec:ablation}

To understand the contribution of each component of \tech{}, we conduct a comprehensive ablation study on MCP-SB and MCPTox, with the results summarized in~\autoref{tab:ablation}.
We also ablate the remaining auxiliary components---token ranking, role-based masking, feedback signals, and optimization traces---and analyze the sensitivity to key hyperparameters, confirming that each design choice contributes meaningfully to \tech{}'s effectiveness (see~\autoref{app:impact_component}).

\pbold{Impact of Localized Token Masking and Injection.}
Disabling the masking mechanism causes one of the largest single-component performance drops across both benchmarks, reducing ASR from 100.0\% to 61.1\% ($\Delta = -38.9$) on MCP-SB and from 88.0\% to 60.0\% ($\Delta = -28.0$) on MCPTox. These results demonstrate that localized token masking and injection effectively guide the optimizer in identifying optimal positions to embed the malicious payload while preserving surrounding context.

\pbold{Impact of Multiple Starting Points.}
Restricting optimization to a single initial candidate (Single Start Point $N=1$) severely degrades ASR from 100.0\% to 61.1\% ($\Delta = -38.9$) on MCP-SB and from 88.0\% to 68.0\% ($\Delta = -20.0$) on MCPTox. Because discrete prompt optimization is prone to local optima when the initial description mutation does not align with the victim model's tool-calling behavior, maintaining multiple starting points ($N > 1$) substantially improves exploration.

\pbold{Impact of Tailored Mutation Instruction.}
Replacing our tailored mutation instruction with a generic rewriting instruction (Naive Mutation Prompt) reduces ASR to 88.9\% ($\Delta = -11.1$) on MCP-SB and causes a sharp drop to 66.7\% ($\Delta = -21.3$) on the more complex MCPTox benchmark. Without tailored mutation instructions, the proposer frequently generates overly directive mutations that are less effective at steering argument values, or simply rewrites the tool description in a way that compromises its original functional fidelity.

\pbold{Joint Ablation.}
Based on the single-component ablations above, we identify the masking mechanism, multi-start initialization, and tailored mutation instructions as the three most impactful components of \tech{}, and further evaluate their combined removal. Removing both the masking mechanism and multi-start search (No Masking + $N=1$) reduces ASR to 50.0\% ($\Delta = -50.0$) on MCP-SB and 64.0\% ($\Delta = -24.0$) on MCPTox. Additionally replacing the tailored mutation instructions (No Masking + $N=1$ + Naive prompt) further lowers ASR to 38.9\% ($\Delta = -61.1$) and 56.0\% ($\Delta = -32.0$), respectively. Finally, removing all \tech{} components (Simple Rewriting) reduces ASR to 16.7\% ($\Delta = -83.3$) on MCP-SB and 12.0\% ($\Delta = -76.0$) on MCPTox, demonstrating that localized masking, multi-start search, and tailored mutation instructions act synergistically rather than redundantly.

\section{Related Work}

\pbold{Prompt Injection Attack}
The goal of prompt injection (PI) is to use malicious instructions to make the agent produce adversarial content~\cite{perez2022ignore, toyer2024tensor} or to hijack the agent's workflow to carry out malicious actions~\cite{debenedetti2024agentdojo, zhan2024injecagent}.
PI can be divided into direct prompt injection (DPI)~\cite{perez2022ignore} and indirect prompt injection (IPI)~\cite{greshake2023not}. The difference between DPI and IPI lies in the channel through which the malicious instruction reaches the agent. In DPI, the instruction is injected directly through the user input. Here the attacker is typically the user, and the goal is to extract the system prompt or the reasoning trace of the LLM~\cite{zhang2024effective, panfilov2026stealing}. In IPI, by contrast, the instruction is injected into an external data source, such as a web page or a document retrieved via retrieval-augmented generation (RAG)~\cite{guo2026hidden}, so the user is typically unaware of the attack.

\pbold{Tool-Use Attacks.}
Unlike conventional indirect prompt injection (IPI), which embeds payloads in external content, recent attacks directly target the LLM tool-use pipeline via tool selection or tool poisoning. Tool-selection attacks induce agents to pick a malicious tool over benign alternatives: \textit{Shi et al.}~\cite{shi2026toolhijacker} and \textit{Mo et al.}~\cite{mo2025attractive} optimize tool descriptions to bias selection, whereas \textit{Wang et al.}~\cite{wang2026mpma} apply advertising-style persuasive rewriting.

Tool poisoning corrupts tool metadata to hijack workflows or tamper with arguments~\cite{invariant2025mcp, zhang2026mcp, zong2026mcp, wang2026mcptox}. Early methods rely on explicit directives such as \texttt{<IMPORTANT>} tags~\cite{invariant2025mcp} or imperative instructions~\cite{wang2026mcptox}, while \textit{Li et al.}~\cite{li2026mcp} propose MCP-ITP, which optimizes descriptions using dual LLM judges for attack success and stealthiness. However, our experiments show that MCP-ITP remains vulnerable to existing defenses because it overlooks the structural characteristics of tool descriptions. In contrast, \tech{} exploits tool-specification properties to achieve high attack effectiveness while evading defenses.

\pbold{Tool-Use Defenses.}
Specialized defenses against tool-use attacks span static scanners, runtime firewalls, and system-level MCP frameworks. Static filter-based scanners include Snyk Agent Scan~\cite{snyk2026agentscan}, Cisco MCP Scan~\cite{cisco2025mcpscanner}, and MCP-Shield~\cite{risen2025mcpshield}, whereas firewall-based defenses~\cite{bhagwatkar2025indirect} employ an LLM judge at runtime to verify that generated tool arguments align with user intent. System-level MCP frameworks further incorporate multi-stage inspection pipelines, such as MCP-Guard~\cite{xing2025mcpguard} and MCP Guardian~\cite{kumar2025mcpguardian}, or cryptographic schema verification~\cite{bhatt2025etdi}.

Existing studies also benchmark conventional IPI defenses; for example, ToolHijacker~\cite{shi2026toolhijacker} evaluates perplexity-based filters~\cite{alon2023detecting, jain2023baseline} and alignment-based training~\cite{chen2025struq, chen2025secalign}. In this work, we evaluate \tech{} against both MCP-specific mechanisms and representative IPI defenses~\cite{schulhoff2023sandwich, zhu2025melon}.

\section{Conclusion}

We present \tech{}, an LLM-based method for poisoning tool descriptions to manipulate tool selection and argument generation.
Using argument-relevant masking and dynamically refined mutation instructions, \tech{} produces concise and natural-appearing descriptions without detector feedback.
Evaluations on three public datasets show higher attack success and lower detectability than existing baselines.
These findings highlight the need to secure tool metadata that directly influences agent execution.

\bibliographystyle{IEEEtran}
\bibliography{main}

@misc{protectai2024deberta,
  author       = {{ProtectAI}},
  title        = {Fine-tuned DeBERTa-v3-base for Prompt Injection Detection},
  year         = {2024},
  howpublished = {\url{https://huggingface.co/ProtectAI/deberta-v3-base-prompt-injection-v2}},
  note         = {Hugging Face model repository}
}

@article{debenedetti2024agentdojo,
  title={Agentdojo: A dynamic environment to evaluate prompt injection attacks and defenses for llm agents},
  author={Debenedetti, Edoardo and Zhang, Jie and Balunovic, Mislav and Beurer-Kellner, Luca and Fischer, Marc and Tram{\`e}r, Florian},
  journal={Advances in neural information processing systems},
  volume={37},
  pages={82895--82920},
  year={2024}
}

@inproceedings{zhan2024injecagent,
  title={Injecagent: Benchmarking indirect prompt injections in tool-integrated large language model agents},
  author={Zhan, Qiusi and Liang, Zhixiang and Ying, Zifan and Kang, Daniel},
  booktitle={Findings of the Association for Computational Linguistics: ACL 2024},
  pages={10471--10506},
  year={2024}
}

@inproceedings{toyer2024tensor,
  title={Tensor trust: Interpretable prompt injection attacks from an online game},
  author={Toyer, Sam and Watkins, Olivia and Mendes, Ethan and Svegliato, Justin and Bailey, Luke and Wang, Tiffany and Ong, Isaac and Elmaaroufi, Karim and Abbeel, Pieter and Ritter, Alan and others},
  booktitle={International Conference on Learning Representations},
  volume={2024},
  pages={18714--18746},
  year={2024}
}

@article{perez2022ignore,
  title={Ignore previous prompt: Attack techniques for language models},
  author={Perez, F{\'a}bio and Ribeiro, Ian},
  journal={arXiv preprint arXiv:2211.09527},
  year={2022}
}

@inproceedings{greshake2023not,
  title={Not what you've signed up for: Compromising real-world llm-integrated applications with indirect prompt injection},
  author={Greshake, Kai and Abdelnabi, Sahar and Mishra, Shailesh and Endres, Christoph and Holz, Thorsten and Fritz, Mario},
  booktitle={Proceedings of the 16th ACM workshop on artificial intelligence and security},
  pages={79--90},
  year={2023}
}

@inproceedings{
zhang2024effective,
title={Effective Prompt Extraction from Language Models},
author={Yiming Zhang and Nicholas Carlini and Daphne Ippolito},
booktitle={First Conference on Language Modeling},
year={2024},
url={https://openreview.net/forum?id=0o95CVdNuz}
}

@article{panfilov2026stealing,
  title={Stealing reasoning traces from proprietary LLM APIs},
  author={Panfilov, Alexander and Schmotz, David and Shumailov, Ilia and Beurer-Kellner, Luca and Schaeffer, Joachim and Prabhu, Ameya and Geiping, Jonas and Andriushchenko, Maksym},
  journal={arXiv preprint arXiv:2608.09867},
  year={2026}
}

@inproceedings{guo2026hidden,
  title={Hidden-in-plain-text: A benchmark for social-web indirect prompt injection in rag},
  author={Guo, Haoze and Wei, Ziqi},
  booktitle={Proceedings of the ACM Web Conference 2026},
  pages={8337--8340},
  year={2026}
}

@article{chennabasappa2025llamafirewall,
  title={Llamafirewall: An open source guardrail system for building secure ai agents},
  author={Chennabasappa, Sahana and Nikolaidis, Cyrus and Song, Daniel and Molnar, David and Ding, Stephanie and Wan, Shengye and Whitman, Spencer and Deason, Lauren and Doucette, Nicholas and Montilla, Abraham and others},
  journal={arXiv preprint arXiv:2505.03574},
  year={2025}
}

@article{zhou2026mcpshield,
  title={Mcpshield: A security cognition layer for adaptive trust calibration in model context protocol agents},
  author={Zhou, Zhenhong and Zhang, Yuanhe and Cai, Hongwei and Aloqaily, Moayad and Bouachir, Ouns and Pang, Linsey and Mehrotra, Prakhar and Wang, Kun and Wen, Qingsong},
  journal={arXiv preprint arXiv:2602.14281},
  year={2026}
}

@inproceedings{liu2025datasentinel,
  title={Datasentinel: A game-theoretic detection of prompt injection attacks},
  author={Liu, Yupei and Jia, Yuqi and Jia, Jinyuan and Song, Dawn and Gong, Neil Zhenqiang},
  booktitle={2025 IEEE Symposium on Security and Privacy (SP)},
  pages={2190--2208},
  year={2025},
  organization={IEEE}
}

@inproceedings{chen2025secalign,
  title={Secalign: Defending against prompt injection with preference optimization},
  author={Chen, Sizhe and Zharmagambetov, Arman and Mahloujifar, Saeed and Chaudhuri, Kamalika and Wagner, David and Guo, Chuan},
  booktitle={Proceedings of the 2025 ACM SIGSAC Conference on Computer and Communications Security},
  pages={2833--2847},
  year={2025}
}

@article{zhu2025melon,
  title={MELON: Provable defense against indirect prompt injection attacks in AI agents},
  author={Zhu, Kaijie and Yang, Xianjun and Wang, Jindong and Guo, Wenbo and Wang, William Yang},
  journal={arXiv preprint arXiv:2502.05174},
  year={2025}
}

@inproceedings{
bhagwatkar2025indirect,
title={Indirect Prompt Injections: Are Firewalls All You Need, or Stronger Benchmarks?},
author={Rishika Bhagwatkar and Kevin Kasa and Abhay Puri and Gabriel Huang and Irina Rish and Graham W. Taylor and Krishnamurthy Dj Dvijotham and Alexandre Lacoste},
booktitle={Lock-LLM Workshop: Prevent Unauthorized Knowledge Use from Large Language Models},
year={2025},
url={https://openreview.net/forum?id=aSUHAayPml}
}

@inproceedings{
shi2026think,
title={Think Twice Before You Act: Protecting {LLM} Agents Against Tool Description Poisoning via Isolated Planning},
author={Shanghao Shi and Xiao Wang and Chaoyu Zhang and Hao Li and Wenjing Lou and Thomas Hou and Yevgeniy Vorobeychik and Chongjie Zhang and Ning Zhang},
booktitle={Forty-third International Conference on Machine Learning},
year={2026},
url={https://openreview.net/forum?id=jiNw5AgBbw}
}

@inproceedings{mo2025attractive,
  title     = {Attractive Metadata Attack: Inducing LLM Agents to Invoke Malicious Tools},
  author    = {Mo, Kanghua and Hu, Li and Long, Yucheng and Li, Zhihao},
  booktitle = {Advances in Neural Information Processing Systems},
  volume    = {38},
  year      = {2025},
  url       = {https://arxiv.org/abs/2508.02110}
}

@inproceedings{wang2026mcptox,
  title     = {{MCPTox}: A Benchmark for Tool Poisoning Attack on Real-World MCP Servers},
  author    = {Wang, Zhiqiang and Gao, Yichao and Tie, Yanting and Liu, Suyuan and Sun, Haifeng and Cheng, Haoran and Shi, Guanquan and Du, Haohua and Li, Xiang-Yang},
  booktitle = {Proceedings of the AAAI Conference on Artificial Intelligence},
  year      = {2026},
  doi       = {10.1609/aaai.v40i42.40895}
}

@inproceedings{chen2025struq,
  title={$\{$StruQ$\}$: Defending against prompt injection with structured queries},
  author={Chen, Sizhe and Piet, Julien and Sitawarin, Chawin and Wagner, David},
  booktitle={34th USENIX Security Symposium (USENIX Security 25)},
  pages={2383--2400},
  year={2025}
}

@article{hasan2025model,
  title={Model context protocol (mcp) at first glance: Studying the security and maintainability of mcp servers},
  author={Hasan, Mohammed Mehedi and Li, Hao and Fallahzadeh, Emad and Rajbahadur, Gopi Krishnan and Adams, Bram and Hassan, Ahmed E},
  journal={arXiv preprint arXiv:2506.13538},
  year={2025}
}

@online{anthropic2024mcp,
  title   = {Introducing the Model Context Protocol},
  author  = {Anthropic},
  year    = {2024},
  note     = {https://www.anthropic.com/news/model-context-protocol}
}

@misc{mcpregistry2024mcp,
  author       = {{Model Context Protocol}},
  title        = {{Official MCP Registry}},
  howpublished = {\url{https://registry.modelcontextprotocol.io/}}
}

@online{mcpmarket2024mcp,
  title   = {MCP Market},
  author  = {MCP Market},
  year    = {2024},
  note     = {https://mcpmarket.com}
}

@online{mcpservers2024mcp,
  title   = {MCP Servers},
  author  = {MCP Servers},
  year    = {2024},
  note     = {https://mcpservers.org}
}

@inproceedings{
yin2026pismith,
title={{PIS}mith: Reinforcement Learning-based Red Teaming for Prompt Injection Defenses},
author={Chenlong Yin and Runpeng Geng and Yanting Wang and Jinyuan Jia},
booktitle={Third Conference on Language Modeling},
year={2026},
url={https://openreview.net/forum?id=tBmgvHPMVT}
}

@inproceedings{nasr2026attacker,
  title={The Attacker Moves Second: Stronger Adaptive Attacks Bypass Defenses Against $\{$LLM$\}$ Jailbreaks and Prompt Injections},
  author={Nasr, Milad and Carlini, Nicholas and Sitawarin, Chawin and Schulhoff, Sander V and Hayes, Jamie and Ilie, Michael and Pluto, Juliette and Song, Shuang and Chaudhari, Harsh and Shumailov, Ilia and others},
  booktitle={35th USENIX Security Symposium (USENIX Security 26)},
  pages={1467--1486},
  year={2026}
}

@article{li2026mcp,
  title={Mcp-itp: An automated framework for implicit tool poisoning in mcp},
  author={Li, Ruiqi and Wang, Zhiqiang and Yao, Yunhao and Li, Xiang-Yang},
  journal={arXiv preprint arXiv:2601.07395},
  year={2026}
}

@online{invariant2025mcp,
  title   = {MCP Security Notification: Tool Poisoning Attacks},
  author  = {{Invariant Labs}},
  year    = {2025},
  date    = {2025-04-01},
  note     = {https://invariantlabs.ai/blog/mcp-security-notification-tool-poisoning-attacks}
}

@inproceedings{zhang2026mcp,
  title={Mcp security bench (msb): Benchmarking attacks against model context protocol in llm agents},
  author={Zhang, Dongsen and Li, Zekun and Luo, Xu and Liu, Xuannan and Li, Pei and Xu, Wenjun},
  booktitle={International Conference on Learning Representations},
  volume={2026},
  pages={58341--58372},
  year={2026}
}

@inproceedings{zong2026mcp,
  title={MCP-safetybench: A benchmark for safety evaluation of large language models with real-world MCP servers},
  author={Zong, Xuanjun and Shen, Zhiqi and Wang, Lei and Lan, Yunshi and others},
  booktitle={International Conference on Learning Representations},
  volume={2026},
  pages={130709--130731},
  year={2026}
}

@article{wang2025promptsleuth,
  title={Promptsleuth: Detecting prompt injection via semantic intent invariance},
  author={Wang, Mengxiao and Zhang, Yuxuan and Gu, Guofei},
  journal={arXiv preprint arXiv:2508.20890},
  year={2025}
}

@inproceedings{schick2023toolformer,
  author    = {Schick, Timo and Dwivedi-Yu, Jane and Dess{\`i}, Roberto and Raileanu, Roberta and Lomeli, Maria and Hambro, Eric and Zettlemoyer, Luke and Cancedda, Nicola and Scialom, Thomas},
  title     = {Toolformer: Language Models Can Teach Themselves to Use Tools},
  booktitle = {Advances in Neural Information Processing Systems (NeurIPS)},
  volume    = {36},
  year      = {2023}
}

@article{wang2024survey,
  author  = {Wang, Lei and Ma, Chen and Feng, Xueyang and Zhang, Zeyu and Yang, Hao and Zhang, Jingsen and Chen, Zhiyuan and Tang, Jiakai and Chen, Xu and Lin, Yankai and Zhao, Wayne Xin and Wei, Zhewei and Wen, Ji-Rong},
  title   = {A Survey on Large Language Model based Autonomous Agents},
  journal = {Frontiers of Computer Science},
  volume  = {18},
  number  = {6},
  pages   = {186345},
  year    = {2024}
}

@article{wallace2024instruction,
  author  = {Wallace, Eric and Xiao, Kai and Leike, Reimar and Weng, Lilian and Heidecke, Johannes and Beutel, Alex},
  title   = {The Instruction Hierarchy: Training {LLMs} to Prioritize Privileged Instructions},
  journal = {arXiv preprint arXiv:2404.13208},
  year    = {2024}
}

@inproceedings{lou2024discrete,
  title     = {Discrete Diffusion Modeling by Estimating the Ratios of the Data Distribution},
  author    = {Lou, Aaron and Meng, Chenlin and Ermon, Stefano},
  booktitle = {International Conference on Machine Learning (ICML)},
  year      = {2024}
}

@inproceedings{sahoo2024simple,
  title     = {Simple and Effective Masked Diffusion Language Models},
  author    = {Sahoo, Subham Sekhar and Arriola, Marianne and Schiff, Yair and Gokaslan, Aaron and Marroquin, Edgar and Chiu, Justin T. and Rush, Alexander and Kuleshov, Volodymyr},
  booktitle = {Advances in Neural Information Processing Systems (NeurIPS)},
  year      = {2024}
}

@inproceedings{nie2025large,
  title     = {Large Language Diffusion Models},
  author    = {Nie, Shen and Zhu, Fengqi and You, Zebin and Zhang, Xiaolu and Ou, Jingyang and Hu, Jun and Zhou, Jun and Lin, Yankai and Wen, Ji-Rong and Li, Chongxuan},
  booktitle = {Advances in Neural Information Processing Systems (NeurIPS)},
  year      = {2025}
}

@article{yin2026climbing,
  title={Climbing the Hill: Prompt Injection Red-Teaming Against Frontier Models with Curriculum Reinforcement Learning},
  author={Yin, Chenlong and Jin, Xiaolong and Zou, Wei and Wang, Yanting and Jia, Jinyuan},
  journal={arXiv preprint arXiv:2609.33628},
  year={2026}
}

@inproceedings{kwon2023efficient,
  title={Efficient Memory Management for Large Language Model Serving with PagedAttention},
  author={Kwon, Woosuk and Li, Zhuohan and Zhuang, Siyuan and Sheng, Ying and Zheng, Lianmin and Yu, Cody Hao and Gonzalez, Joseph E. and Zhang, Hao and Stoica, Ion},
  booktitle={Proceedings of the 29th Symposium on Operating Systems Principles},
  pages={611--626},
  year={2023}
}

@misc{cisco2025mcpscanner,
  author       = {{Cisco AI Defense}},
  title        = {{MCP Scanner}: A Python Tool for Scanning {MCP} Servers and Tools for Potential Security Findings},
  year         = {2025},
  howpublished = {\url{https://github.com/cisco-ai-defense/mcp-scanner}},
  note         = {Accessed: 2026}
}

@misc{risen2025mcpshield,
  title   = {MCP-Shield},
  author  = {Nikita Kryzhanouski},
  year    = {2025},
  note     = {https://github.com/riseandignite/mcp-shield}
}

@misc{snyk2026agentscan,
  author       = {{Snyk}},
  title        = {{Agent Scan}: Security Scanner for {AI} Agents, {MCP} Servers and Agent Skills},
  year         = {2026},
  howpublished = {\url{https://github.com/snyk/agent-scan}}
}

@misc{schulhoff2023sandwich,
  author       = {Schulhoff, Sander},
  title        = {The Sandwich Defense: Strengthening {AI} Prompt Security},
  year         = {2023},
  howpublished = {\url{https://learnprompting.org/docs/prompt_hacking/defensive_measures/sandwich_defense}},
  note         = {Learn Prompting}
}

@inproceedings{liu2024formalizing,
  author    = {Liu, Yupei and Jia, Yuqi and Geng, Runpeng and Jia, Jinyuan and Gong, Neil Zhenqiang},
  title     = {Formalizing and Benchmarking Prompt Injection Attacks and Defenses},
  booktitle = {33rd USENIX Security Symposium (USENIX Security 24)},
  year      = {2024}
}

@misc{qwen38,
  title        = {{Qwen3.8-Max}: A New Bar for Coding and Cowork},
  author       = {{Qwen Team}},
  year         = {2026},
  month        = {August},
  howpublished = {\url{https://qwen.ai/blog?id=qwen3.8}},
  url          = {https://qwen.ai/blog?id=qwen3.8}
}

@article{rafailov2023direct,
  title={Direct preference optimization: Your language model is secretly a reward model},
  author={Rafailov, Rafael and Sharma, Archit and Mitchell, Eric and Manning, Christopher D and Ermon, Stefano and Finn, Chelsea},
  journal={Advances in neural information processing systems},
  volume={36},
  pages={53728--53741},
  year={2023}
}

@article{hendrycks2020measuring,
  title={Measuring massive multitask language understanding},
  author={Hendrycks, Dan and Burns, Collin and Basart, Steven and Zou, Andy and Mazeika, Mantas and Song, Dawn and Steinhardt, Jacob},
  journal={arXiv preprint arXiv:2009.03300},
  year={2020}
}

@misc{webmcp2026sec,
  author       = {{Chrome for Developers}},
  title        = {{WebMCP} Tool Security: Set Character Budgets},
  year         = {2026},
  howpublished = {\url{https://developer.chrome.com/docs/ai/webmcp/secure-tools}}
}

@techreport{csa2026mcp,
  author      = {{Cloud Security Alliance}},
  title       = {{MCP} Tool Poisoning: Adversarial Hijacking of {AI} Agent Workflows},
  institution = {Cloud Security Alliance},
  year        = {2026},
  url         = {https://labs.cloudsecurityalliance.org/research/csa-research-note-mcp-tool-poisoning-ai-agent-exfiltration-2/}
}

@misc{llamaindex-lengthy-tools,
  author       = {{LlamaIndex}},
  title        = {{OpenAI} Agent Workarounds for Lengthy Tool Descriptions},
  year         = {2024},
  howpublished = {\url{https://developers.llamaindex.ai/python/examples/agent/openai_agent_lengthy_tools/}},
}

@misc{anthropic-mcp-desc-length,
  author       = {{Anthropic}},
  title        = {Connect {Claude Code} to Tools via {MCP}},
  year         = {2026},
  howpublished = {\url{https://code.claude.com/docs/en/mcp}},
}

@inproceedings{patil2024gorilla,
  author    = {Patil, Shishir G. and Zhang, Tianjun and Wang, Xin and Gonzalez, Joseph E.},
  title     = {Gorilla: Large Language Model Connected with Massive {APIs}},
  booktitle = {Advances in Neural Information Processing Systems (NeurIPS)},
  volume    = {37},
  pages     = {126544--126565},
  year      = {2024}
}

@inproceedings{reimers2019sentencebert,
  author    = {Reimers, Nils and Gurevych, Iryna},
  title     = {Sentence-{BERT}: Sentence Embeddings using Siamese {BERT}-Networks},
  booktitle = {Proceedings of the 2019 Conference on Empirical Methods in Natural Language Processing and the 9th International Joint Conference on Natural Language Processing (EMNLP-IJCNLP)},
  pages     = {3982--3992},
  year      = {2019},
  publisher = {Association for Computational Linguistics}
}

@inproceedings{wang2020minilm,
  author    = {Wang, Wenhui and Wei, Furu and Dong, Li and Bao, Hangbo and Yang, Nan and Zhou, Ming},
  title     = {{MiniLM}: Deep Self-Attention Distillation for Task-Agnostic Compression of Pre-Trained Transformers},
  booktitle = {Advances in Neural Information Processing Systems (NeurIPS)},
  volume    = {33},
  pages     = {5776--5788},
  year      = {2020}
}

@inproceedings{shi2026toolhijacker,
  author    = {Shi, Jiawen and Yuan, Zenghui and Tie, Guiyao and Zhou, Pan and Gong, Neil Zhenqiang and Sun, Lichao},
  title     = {Prompt Injection Attack to Tool Selection in {LLM} Agents},
  booktitle = {Proceedings of the Network and Distributed System Security (NDSS) Symposium},
  year      = {2026}
}

@inproceedings{wang2026mpma,
  author    = {Wang, Zihan and Zhang, Rui and Liu, Yu and Fan, Wenshu and Jiang, Wenbo and Zhao, Qingchuan and Li, Hongwei and Xu, Guowen},
  title     = {{MPMA}: Preference Manipulation Attack Against Model Context Protocol},
  booktitle = {Proceedings of the 40th Annual AAAI Conference on Artificial Intelligence (AAAI)},
  volume    = {40},
  number    = {42},
  pages     = {35838--35846},
  year      = {2026}
}

@article{alon2023detecting,
  author  = {Alon, Gabriel and Kamfonas, Michael},
  title   = {Detecting Language Model Attacks with Perplexity},
  journal = {arXiv preprint arXiv:2308.14132},
  year    = {2023}
}

@article{jain2023baseline,
  author  = {Jain, Neel and Schwarzschild, Avi and Wen, Yuxin and Somepalli, Gowthami and Kirchenbauer, John and Chiang, Ping-yeh and Goldblum, Micah and Saha, Aniruddha and Geiping, Jonas and Goldstein, Tom},
  title   = {Baseline Defenses for Adversarial Attacks Against Aligned Language Models},
  journal = {arXiv preprint arXiv:2309.00614},
  year    = {2023}
}

@article{xing2025mcpguard,
  author  = {Xing, Wenpeng and Li, Zhonghao and Lin, Changting and Han, Meng},
  title   = {{MCP-Guard}: A Defense Framework for Model Context Protocol Integrity in Large Language Model Applications},
  journal = {arXiv preprint arXiv:2508.10991},
  year    = {2025}
}

@article{kumar2025mcpguardian,
  author  = {Kumar, Sonu and Girdhar, Anubhav and Patil, Ritesh and Tripathi, Divyanshu},
  title   = {{MCP Guardian}: A Security-First Layer for Safeguarding {MCP}-Based {AI} System},
  journal = {arXiv preprint arXiv:2504.12757},
  year    = {2025}
}

@article{bhatt2025etdi,
  author  = {Bhatt, Manish and Narajala, Vineeth Sai and Habler, Idan},
  title   = {{ETDI}: Mitigating Tool Squatting and Rug Pull Attacks in Model Context Protocol by Using {OAuth}-Enhanced Tool Definitions and Policy-Based Access Control},
  journal = {arXiv preprint arXiv:2506.13538},
  year    = {2025}
}

@article{hu2021lora,
  title={Lora: Low-rank adaptation of large language models},
  author={Hu, Edward J and Shen, Yelong and Wallis, Phillip and Allen-Zhu, Zeyuan and Li, Yuanzhi and Wang, Shean and Wang, Lu and Chen, Weizhu},
  journal={arXiv preprint arXiv:2106.09685},
  year={2021}
}

\appendices

\section{Datasets}
\label{app:datasets}

\pbold{MCPTox~\cite{wang2026mcptox}}
is a tool-poisoning benchmark of $84$ poisoned MCP tools, each paired with a malicious
objective that overwrites one argument of a targeted tool call. 
The tools span $23$ real-world MCP servers (e.g.,
filesystem, SQLite, GitHub, Slack, and email), covering $84$ distinct targeted tools.
We keep the $75$ items where the attack changes one existing argument from a clear
benign value to a malicious one, so success is unambiguous to measure.

\pbold{MCP-SafetyBench (MCP-SB)~\cite{zong2026mcp}}
is a parameter-poisoning benchmark of $18$ items over $5$ MCP servers, in which the
poisoned description embeds a \emph{standing directive} (a social-engineering narrative
instructing the agent to silently overwrite an argument). 
Each item has a unique targeted tool and the target argument.

\pbold{MCP Security Bench (MSB)~\cite{zhang2026mcp}}
is a parameter-poisoning benchmark of $16$ items over $5$ MCP servers
with $16$ distinct targeted tools, in which the injection is \emph{parasitic on the user's own tool call}---it adds an out-of-scope parameter (with a fixed malicious value) to a tool the user legitimately invokes. 

\begin{table}[t]
\centering
\caption{Length-matched baseline \textbf{sASR} (\%) Columns:
PISmith (\textsc{N}=nano, \textsc{4o}=4o-mini) and CTH (\textsc{T}=terra, \textsc{L}=luna).
}
\label{tab:lenmatch-sasr}
\scriptsize
\setlength{\tabcolsep}{2.6pt}
\begin{tabular}{l cccc cccc cccc}
\toprule
& \multicolumn{4}{c}{\textbf{MSB}} & \multicolumn{4}{c}{\textbf{MCP-SB}} & \multicolumn{4}{c}{\textbf{MCPTox}} \\
\cmidrule(lr){2-5}\cmidrule(lr){6-9}\cmidrule(lr){10-13}
Victim & N & 4o & T & L & N & 4o & T & L & N & 4o & T & L \\
\midrule
SecAlign-8B   & 12 & 6 & 19 & 25 & 39 &  6 & 22 & 17 &  4 & 0 &  4 &  3 \\
SecAlign-70B  &  6 & 0 &  0 &  6 & 17 &  6 & 28 & 17 &  4 & 0 &  1 &  1 \\
Llama-8B      & 38 & 6 & 31 & 69 & 22 &  6 & 17 & 11 &  1 & 0 &  0 &  0 \\
Llama-70B     & 44 & 6 & 31 & 69 & 33 & 11 & 44 & 39 & 15 & 1 & 16 & 17 \\
Qwen3.8-27B   & 25 & 0 & 12 & 31 & 61 & 11 & 50 & 61 & 44 & 3 & 35 & 55 \\
Qwen3-32B     & 31 & 6 & 19 & 38 & 33 & 11 & 39 & 28 & 17 & 4 &  8 & 17 \\
Muse      & 19 & 0 &  0 & 38 & 33 & 11 & 44 & 39 & 61 & 5 & 41 & 51 \\
GPT5.6-Luna   & 75 & 6 & 81 & 88 & 61 & 11 & 50 & 56 &  5 & 0 &  1 &  8 \\
\midrule
Avg.          & 31.3 & 3.8 & 24.1 & 45.5 & 37.4 & 9.1 & 36.8 & 33.5 & 18.9 & 1.6 & 13.3 & 19.0 \\
\bottomrule
\end{tabular}
\end{table}

\section{Length-Matched Comparison}
\label{app:lenmatch}

As shown in~\autoref{tab:injection-signal}, PISmith and CTH produce substantially longer tool descriptions than all other baselines (307--805 tokens), with CTH-Luna averaging $11.5\times$ the length of \tech{} (70 tokens). In practical deployments, however, agentic service providers frequently enforce maximum schema description lengths~\cite{llamaindex-lengthy-tools,anthropic-mcp-desc-length} or flag anomalously verbose tool metadata during initial inspection~\cite{webmcp2026sec,csa2026mcp}. 
To isolate the impact of description verbosity, we constrain PISmith and CTH during generation to match \tech{}'s token number and compare their length-matched sASR (\autoref{tab:lenmatch-sasr}) against their unconstrained performance (\autoref{tab:msb_results}--\autoref{tab:secalign_all}). 
Without verbose contextual padding, every baseline shows a substantial decline in sASR across the eight victim models on each benchmark.
On MSB (where \tech{} averages 95.3\%), average sASR drops from 72.7\% to 45.5\% for {CTH-Luna}, from 52.3\% to 24.1\% for {CTH-Terra}, and from 56.3\% to 31.3\% for {PISmith-Nano}.
The observations are similar on the other two benchmarks.
Meanwhile, {PISmith-4o} remains largely ineffective across all three benchmarks in both settings (0.0\%--1.5\% unconstrained vs.\ 1.6\%--9.1\% length-matched).
Most notably, matching description lengths sharply reduces baseline effectiveness against models with alignment-based defenses ({SecAlign-8B} and {SecAlign-70B}).
On MSB and MCP-SB, {CTH-Luna} drops from 75.0\% to 15.5\% and 17.0\%, while {CTH-Terra} and {PISmith-Nano} fall to 9.0\%--28.0\% (compared to 100.0\% and 80.6\% for \tech{}).
On MCPTox all four length-matched baselines degrade to 0.0\%--4.0\% (vs.\ 57.8\% for \tech{}). 
These results show that existing RL-based baselines rely on long and wordy descriptions to hide malicious instructions and bypass model defenses, whereas \tech{} succeeds even within a short 70-token description.

\begin{table}[t]
\centering
\caption{Lexical injection signal in the poisoned tool descriptions.}
\label{tab:injection-signal}
\scriptsize
\begin{tabular}{l r r r r}
\toprule
Method & Tokens & Override $\downarrow$ & Imp-density $\downarrow$& decl/imp $\uparrow$ \\
\midrule
\textbf{GRAFT} & \textbf{70} & \textbf{0.3} & \textbf{0.6} & \textbf{11.7} \\
Combined       & 109         & 4.0          & 4.9          & 1.1  \\
Enhanced-1     & 94          & 1.0          & 3.3          & 1.5  \\
Enhanced-2     & 95          & 4.0          & 5.7          & 1.1  \\
AttackMove     & 85          & 1.0          & 3.5          & 1.5  \\
MCP-ITP        & 106         & 1.0          & 4.0          & 1.5  \\
PISmith-Nano   & 337         & 2.3          & 3.4          & 1.0  \\
PISmith-4o     & 307         & 1.4          & 0.9          & 4.0  \\
CTH-Terra      & 465         & 0.6          & 1.3          & 2.9  \\
CTH-Luna       & 805         & 1.1          & 1.9          & 2.3  \\
\midrule
$r$ with DET@5   & $-0.27$ & $+0.67$ & $+0.59$ & $-0.59$ \\
$r$ with sASR    & $+0.27$ & $-0.67$ & $-0.58$ & $+0.62$ \\
\bottomrule
\end{tabular}

\end{table}

\section{Qualitative Analysis}
\label{sec:qualitative}

To quantify why poisoned tool descriptions trigger or evade static detectors, we analyze three linguistic features alongside token length: (1)~\textbf{override prevalence}, the average number of explicit control or hijacking keywords (e.g., \emph{ignore}, \emph{priority}, \emph{enforce}) per description; (2)~\textbf{imperative density}, the frequency of commanding verbs and directive constructions per 100 tokens, measuring how overtly an attack orders the agent to act; and (3)~\textbf{declarative-to-imperative ratio (decl/imp ratio)}, the ratio of descriptive, specification-style statements to imperative commands, capturing the extent to which an attack legitimizes the injected parameter as a standard functional requirement rather than issuing a direct order. \autoref{tab:injection-signal} shows the result.

Across these metrics, \tech{} exhibits a fundamentally different linguistic profile from all competing baselines. A \tech{} description averages only 70 tokens, contains virtually no override language (0.3 override terms on average), and achieves the lowest imperative density of any method (0.6 per 100 tokens). Crucially, \tech{} \emph{legitimizes} the attacker's argument rather than \emph{commanding} it, yielding a decl/imp ratio of $11.7$ compared to at most $4.0$ for every competing attack. 
Both template-based attacks (Combined, Enhanced) and optimization/RL-based attacks (AttackMove, MCP-ITP, PISmith, CTH) retain the original benign text and append an explicit command, carrying up to $4.0$ override terms and an imperative density of $3.3$--$5.7$ per 100 tokens.

The bottom two rows report the Pearson correlation coefficient $r \in [-1, +1]$ between each linguistic column and the outcome (DET@5 or sASR) across the $n = 10$ methods, where positive values indicate that the feature and outcome increase together, negative values indicate inverse movement, and $|r|$ measures linear strength ($|r| = 1$ denotes a perfect linear relationship and $0$ denotes none). Given the small sample size ($n = 10$), we interpret $|r| \approx 0.6$ as \emph{moderate, directional} evidence rather than a precise effect size. Override prevalence is the strongest predictor of detection ($r = +0.67$) and of {reduced} stealthy attack success ($r = -0.67$).
The decl/imp ratio is the strongest predictor of detector-survived success ($r = +0.62$) and of evading detection ($r = -0.59$).
Description length and imperative density exhibit the same directional signs with weaker magnitudes. In short, \tech{} succeeds not by crafting a stronger injection, but by \emph{not reading as an injection at all}.
It poisons the tool specification using concise and declarative language that presents malicious behavior as legitimate tool functionality.
This framing supports attack success while avoiding the imperative language and explicit override instructions that detectors target.

\section{Adversarial DPO Fine-Tuning}
\label{app:adaptive_dpo}
To evaluate a learning-based defense alongside runtime mitigations, we fine-tune the victim model (\texttt{Qwen/Qwen3.8-27B}) using Direct Preference Optimization (DPO)~\cite{rafailov2023direct} on adversarial examples generated by GRAFT. Specifically, we train a parameter-efficient LoRA adapter~\cite{hu2021lora} (rank $r = 16$) with DPO regularization coefficient $\beta = 0.1$, while keeping the base checkpoint frozen both as the DPO reference policy and as GRAFT's attack optimizer. All preference pairs are constructed exclusively from the training split, with held-out evaluation items strictly isolated and contributing zero training pairs.

Training proceeds in two stages to balance attack resistance against benign tool-use utility:
\begin{enumerate}
    \item \textbf{Resistance Stage:} We construct adversarial preference triples $(x, y_w, y_l)$ from legitimate user queries paired with GRAFT-poisoned tool descriptions. The rejected completion $y_l$ is the poisoned tool call containing attacker-controlled arguments, whereas the chosen completion $y_w$ is the intended clean tool call validated against recovered benign-gold arguments.
    \item \textbf{Utility-Preserving Stage:} To prevent degenerate over-refusal (declining valid requests or emitting empty tool calls on unpoisoned inputs), we augment the training mixture with benign-behavior preference pairs on clean tool descriptions, where $y_w$ is the correct clean tool call and $y_l$ is an invalid or empty (null) output.
\end{enumerate}
Across adapters, we evaluate both held-out attack success rate and clean tool-selection accuracy, and verify general-purpose reasoning stability using MMLU~\cite{hendrycks2020measuring}.

\section{Impact of Components}
\label{app:impact_component}
\pbold{Auxiliary components.}
We also ablate the remaining auxiliary components---token ranking, role-based masking, feedback signal, and optimization traces---and find that each contributes meaningfully to \tech{}'s effectiveness (See~\autoref{tab:ablation}).
Removing target-relevance token ranking (No Token Ranking) lowers ASR to 88.9\% ($\Delta = -11.1$) on MCP-SB and 72.0\% ($\Delta = -16.0$) on MCPTox, as unranked masking fails to prioritize high-leverage semantic and syntactic tokens governing parameter constraints and thus weakens structural payload integration. 
Similarly, ablating role-aware token selection (No Role-based Mask) reduces ASR to 94.4\% ($\Delta = -5.6$) on MCP-SB and 76.0\% ($\Delta = -12.0$) on MCPTox. 
On the optimization side, replacing fine-grained 5-way execution feedback with coarse binary pass/fail signals (Binary Feedback) reduces ASR to 88.9\% ($\Delta = -11.1$) on MCP-SB and 80.0\% ($\Delta = -8.0$) on MCPTox.
Because fine-grained feedback distinguishes intermediate failure modes to provide clearer directional guidance, such as ignoring the poisoned tool, invoking it with clean arguments, or partially satisfying the hijacked condition. 
Finally, omitting historical search context (No Recent Traces) lowers ASR to 83.3\% ($\Delta = -16.7$) on MCP-SB and 80.0\% ($\Delta = -8.0$) on MCPTox. 
These results confirm that all four components are essential to maximizing \tech{}'s effectiveness.

\begin{table}[t]
\centering
\caption{Hyperparameter sensitivity analysis (ASR, \%), where $\Delta$ denotes the change relative to \tech{}. The hyperparameter used for the main experiment result of \tech{} is ($k{=}5$, $N{=}7$, iter $100$, \#queries$=5$)}
\label{tab:hyperparam-ablation}
\scriptsize
\begin{tabular}{lrrrr}
\toprule
& \multicolumn{2}{c}{MCP-SB} & \multicolumn{2}{c}{MCPTox} \\
\cmidrule(lr){2-3} \cmidrule(lr){4-5}
Ablation & ASR & $\Delta$ & ASR & $\Delta$ \\
\midrule
\tech{} & \textbf{100.0} & --- & \textbf{88.0} & --- \\
\midrule
$k{=}1$ number of mask & 88.9 & $-11.1$ & 68.0 & $-20.0$ \\
$k{=}3$ & 94.4 & $-5.6$  & 84.0 & $-4.0$  \\
$k{=}7$ & 88.9 & $-11.1$ & 84.0 & $-4.0$  \\
\midrule
$N{=}1$ starting point & 61.1  & $-38.9$ & 68.0 & $-20.0$ \\
$N{=}5$             & 100.0 & $0.0$   & 68.0 & $-20.0$ \\
$N{=}10$            & 94.4  & $-5.6$  & 86.6 & $-1.4$  \\
$N{=}15$            & 100.0 & $0.0$   & 80.0 & $-8.0$  \\
\midrule
Train queries $=1$ & 61.1 & $-38.9$ & 56.0 & $-32.0$ \\
Train queries $=3$ & 88.9 & $-11.1$ & 80.0 & $-8.0$  \\
Train queries $=7$ & 94.4 & $-5.6$  & 88.0 & $0.0$   \\
\midrule
Iterations $=10$  & 83.3  & $-16.7$ & 56.0 & $-32.0$ \\
Iterations $=30$  & 83.3  & $-16.7$ & 72.0 & $-16.0$ \\
Iterations $=50$  & 94.4  & $-5.6$  & 76.0 & $-12.0$ \\
Iterations $=150$ & 100.0 & $0.0$   & 86.6 & $-1.4$  \\
\bottomrule
\end{tabular}
\end{table}
\pbold{Impact of Hyperparameters.}
We further examine the sensitivity of \tech{} to its four key hyperparameters---mask count $k$, number of starting points $N$, number of training queries, and optimization iterations---around our default configuration ($k=5$, $N=7$, $5$ training queries, and $100$ iterations).
The result are reported in \autoref{tab:hyperparam-ablation}. 
Varying the number of masked tokens $k \in \{1, 3, 5, 7\}$ reveals an optimal trade-off at $k=5$: restricting edits to a single mask ($k=1$) limits rewriting flexibility and lowers ASR to 88.9\% ($\Delta = -11.1$) on MCP-SB and 68.0\% ($\Delta = -20.0$) on MCPTox, whereas over-masking ($k=7$) slightly degrades performance (88.9\% and 84.0\%) by disrupting too much of the original tool context. 
For initialization breadth, increasing $N$ from $1$ to $7$ boosts ASR from 61.1\% to 100.0\% on MCP-SB and from 68.0\% to 88.0\% on MCPTox, while larger values ($N \in \{10, 15\}$) saturate or dip slightly (e.g., 80.0\% on MCPTox at $N=15$). The result suggests that employing multiple starting points increases the search space for LLM proposer to explore.
Meanwhile, optimizing on too few user queries causes query overfitting.
Using a single training query drops ASR to 61.1\% ($\Delta = -38.9$) on MCP-SB and 56.0\% ($\Delta = -32.0$) on MCPTox, whereas scaling to $5$ queries achieves strong cross-query generalization that plateaus at $7$ queries (94.4\% and 88.0\%). 
Finally, ASR improves steadily as the optimization budget increases from $10$ iterations (83.3\% on MCP-SB and 56.0\% on MCPTox) to $100$ iterations, after which performance converges (100.0\% and 86.6\% at $150$ iterations), confirming that our default configuration effectively balances attack efficacy and search overhead.\looseness=-1

\section{Overhead Analysis}
\label{app:overhead}
\begin{table}[t]
\centering
\caption{Result of Overhead Analysis.}
\label{tab:cost-vs-sasr}
\scriptsize
\setlength{\tabcolsep}{4.5pt}
\resizebox{\columnwidth}{!}{%
\begin{tabular}{l r r r r}
\toprule
& \multicolumn{2}{c}{\textbf{Cost / Successful Description}} & & \\
\cmidrule(lr){2-3}
\textbf{Method} & \textbf{LLM Calls} & \textbf{Tokens} & \textbf{Avg.\ sASR (\%)} & \shortstack{\textbf{$\Delta$sASR /}\\\textbf{+1k Calls}} \\
\midrule
MCP-ITP  & $\sim$52      & 0.08\,M       & 48.8          & ---  \\
AttackMove         & $\sim$2{,}000 & $\sim$1.90\,M & 68.7          & 10.2 \\
\tech{}   & $\sim$1{,}227 & $\sim$1.23\,M & \textbf{85.8} & \textbf{31.5} \\
\bottomrule
\end{tabular}%
}
\end{table}

To better understand the computational cost of optimizing a poisoned tool description compared to other optimization-based baselines (AttackMove and MCP-ITP), we further evaluate the average overhead---measured in LLM calls and token consumption---required to generate a successful attack across the three benchmarks. 
As shown in~\autoref{tab:cost-vs-sasr}, \tech{} achieves the highest average sASR (85.8\%), substantially outperforming AttackMove (68.6\%) and MCP-ITP (48.8\%). 
In terms of search overhead, MCP-ITP is the least expensive, requiring an average of $52$ LLM calls and $0.08\text{M}$ tokens per successful attack, followed by \tech{} ($1{,}227$ LLM calls and $1.23\text{M}$ tokens), whereas AttackMove is the most expensive ($2{,}000$ LLM calls and $1.9\text{M}$ tokens). 
Because both \tech{} and AttackMove incur higher query budgets than MCP-ITP, a natural question is whether this additional overhead is justified by the resulting sASR improvement. 
To quantify cost-effectiveness, we compute the marginal sASR gain per $1\text{K}$ additional LLM calls relative to MCP-ITP, finding that \tech{} yields a 31.5\% sASR gain per $1\text{K}$ calls---more than triple the efficiency of AttackMove. 
Based on OpenRouter API pricing for Qwen3.8-27B\footnote{\url{https://openrouter.ai/qwen/qwen3.8-27b}} (starting at \$0.035 per 1M input tokens), and given that $1\text{K}$ LLM calls corresponds to roughly $1\text{M}$ tokens in our setup, \tech{} delivers a 31.5\% sASR gain for roughly \$0.035 in marginal token cost.

\section{Reproducibility}
\label{app:reproduce}

To validate that our primary findings do not depend on deterministic greedy decoding, we further evaluate the stability of \tech{} under stochastic generation. 
On MCPTox, \tech{} achieves an 88.0\% held-out ASR under greedy decoding (deterministic across random seeds) and $83.6\% \pm 3.3\%$ across three stochastic runs at temperature $T = 0.7$, %
with every individual run reaching at least 80.0\%. 
Although temperature sampling introduces minor run-to-run variance, \tech{} maintains high attack success ($\ge 80.0\%$) with only a 4.4\% mean reduction relative to greedy decoding ($88.0\% \rightarrow 83.6\%$), demonstrating strong reproducibility and robustness across realistic deployment decoding configurations.\looseness=-1

\end{document}